\documentclass{article}
\usepackage[preprint]{colm2026_conference}
\usepackage[T1]{fontenc}
\usepackage[utf8]{inputenc}
\usepackage{microtype}
\usepackage{graphicx}
\usepackage{amsmath,amssymb,amsfonts}
\usepackage{booktabs,multirow,tabularx,array}
\usepackage{placeins,float,wrapfig}
\usepackage{xcolor}
\usepackage{url}
\usepackage{hyperref}
\usepackage[nameinlink,noabbrev]{cleveref}
\crefname{section}{Section}{Sections}
\crefname{subsection}{Section}{Sections}
\crefname{subsubsection}{Section}{Sections}
\crefname{appendix}{App.}{Apps.}
\definecolor{darkblue}{rgb}{0, 0, 0.5}
\hypersetup{colorlinks=true,linkcolor=darkblue,citecolor=darkblue,urlcolor=darkblue,
    pdftitle={Backdoor in the Loop: Compromising Agentic Search via Malicious Retrievers},
    pdfauthor={}}

\newif\ifcomment
\commentfalse

\ifcomment
    \newcounter{YXNumberOfComments}
    \stepcounter{YXNumberOfComments}
    \DeclareRobustCommand{\xym}[1]{\textcolor{orange}{\small \bf [XYM\#\arabic{YXNumberOfComments}\stepcounter{YXNumberOfComments}: #1]}}

    \newcounter{PHNumberOfComments}
    \stepcounter{PHNumberOfComments}
    \DeclareRobustCommand{\hpc}[1]{\textcolor{blue}{\small \bf [HPC\#\arabic{PHNumberOfComments}\stepcounter{PHNumberOfComments}: #1]}}  
    
    \DeclareRobustCommand{\del}[1]{{{\color{red}\st{#1}}}}
\else
    \DeclareRobustCommand\xym[1]{}    
    \DeclareRobustCommand\hpc[1]{} 
    \DeclareRobustCommand{\del}[1]{}
\fi

\usepackage{cleveref}
\crefname{figure}{Figure}{Figures}
\Crefname{figure}{Figure}{Figures}

\newcolumntype{Y}{>{\centering\arraybackslash}X}
\newcommand{\dou}{DOU}

\begin{document}

\date{}

\title{Backdoor in the Loop: Compromising Agentic Search via Malicious Retrievers}

\author{Beining Xu$^{*1}$ \quad Peichun Hua$^{*2}$ \quad Yunming Xiao$^{2\dagger}$ \\
$^{1}$Shenzhen MSU-BIT University \quad $^{2}$The Chinese University of Hong Kong, Shenzhen \\
\texttt{xubeining88@gmail.com} \quad \texttt{peichunhua@link.cuhk.edu.cn} \quad \texttt{yunmingxiao@cuhk.edu.cn}
}

\maketitle
{\renewcommand{\thefootnote}{\fnsymbol{footnote}}\footnotetext[1]{Equal contribution.}\footnotetext[2]{Corresponding author.}}
\begin{abstract}
Agentic retrieval-augmented generation (RAG) interleaves reasoning with repeated retrieval, giving the retriever influence over both the evidence an agent observes and its subsequent search decisions. We study retriever backdoors that exploit this feedback loop and repurpose weak backdoor purification to conceal their presence. An attacker supplies a compromised retriever checkpoint while leaving the search agent and deployment corpus unchanged. Without corpus write access, the attacker can still suppress useful evidence, persistently retrieve a selected existing document, or steer the agent toward prolonged search, inflating retrieval, context, and latency cost. To conceal these behaviors from detection, we propose leveraging a controlled inject-and-remove cycle: deliberately inject a weaker backdoor and then unlearn it. This process weakens detector-visible signatures and fools the backdoor detectors with an illusion of purification while preserving the malicious retrieval behavior. These findings expose a systematic vulnerability in RAG systems in which a weak defense becomes an attacker's concealment tool for a backdoored retriever, even when the underlying corpus remains trustworthy.
\end{abstract}

\section{Introduction}
\label{sec:introduction}

Retrieval-augmented generation (RAG) has become a widely used approach for grounding large language models in external knowledge~\citep{lewis2020rag}. Agentic RAG extends this approach by interleaving reasoning with search: an agent identifies information gaps, retrieves evidence, and decides whether to search again. Recent work advances both search policies~\citep{li2025searcho1,jin2025searchr1,chen2025research,song2025r1searcher} and retrievers tailored to agentic workflows~\citep{liu2026agenticr,chen2026agentir,hua2026semantics}. This growing capability also expands the retriever's influence: each retrieved passage can shape the final answer, the next query, and the decision to stop.

Prior attacks compromise RAG through both its evidence store and its retriever. Corpus-poisoning methods craft passages that rank highly and steer downstream answers, as demonstrated by adversarial passage injection and PoisonedRAG~\citep{zhong2023poisoning,zou2025poisonedrag}. Retriever attacks support several activation conditions: BaD-DPR uses natural grammatical errors to favor attacker-inserted passages~\citep{long2025baddpr}; Backdoored Retrievers maps queries about a selected topic to a fixed prompt-injection document~\citep{clop2024backdoored}; and TrojanRAG associates discrete or semantic triggers with malicious contexts~\citep{cheng2024trojanrag}. DisarmRAG and CAREAttack target specified victim queries or prompts, promoting bypass instructions or selected passages while preserving retrieval on non-target inputs~\citep{dai2026disarmrag,liu2026careattack}.
Across these approaches, the relevant distinction is \textit{what} activates the attack and \textit{how} the target evidence is selected, irrespective of whether the retriever is poisoned, fine-tuned, or edited.

Agentic RAG, however, changes \textit{who} determines the search queries. The user supplies an initial question, while the agent generates subsequent queries from its evolving reasoning and retrieved evidence~\citep{jin2025searchr1,liu2026agenticr}. In our Agentic-R setting, each retrieval input combines the original question with an agent-generated query. A trigger in the original question thus remains present across rounds, but changing queries can alter document rankings and the evidence returned. Success on a single retrieval request therefore does not guarantee success across the search trajectory. This creates a trajectory-level challenge: the attacker cannot directly control the agent's subsequent queries, reasoning, or stopping decisions. Retrieved passages can reshape the agent's next query and decision to continue searching, so an effective attack must sustain its effect through this feedback loop to cause an incorrect answer, repeated target retrieval, or an inappropriate stopping decision.

\begin{wrapfigure}{r}{0.49\textwidth}
    \centering
    \includegraphics[width=\linewidth]{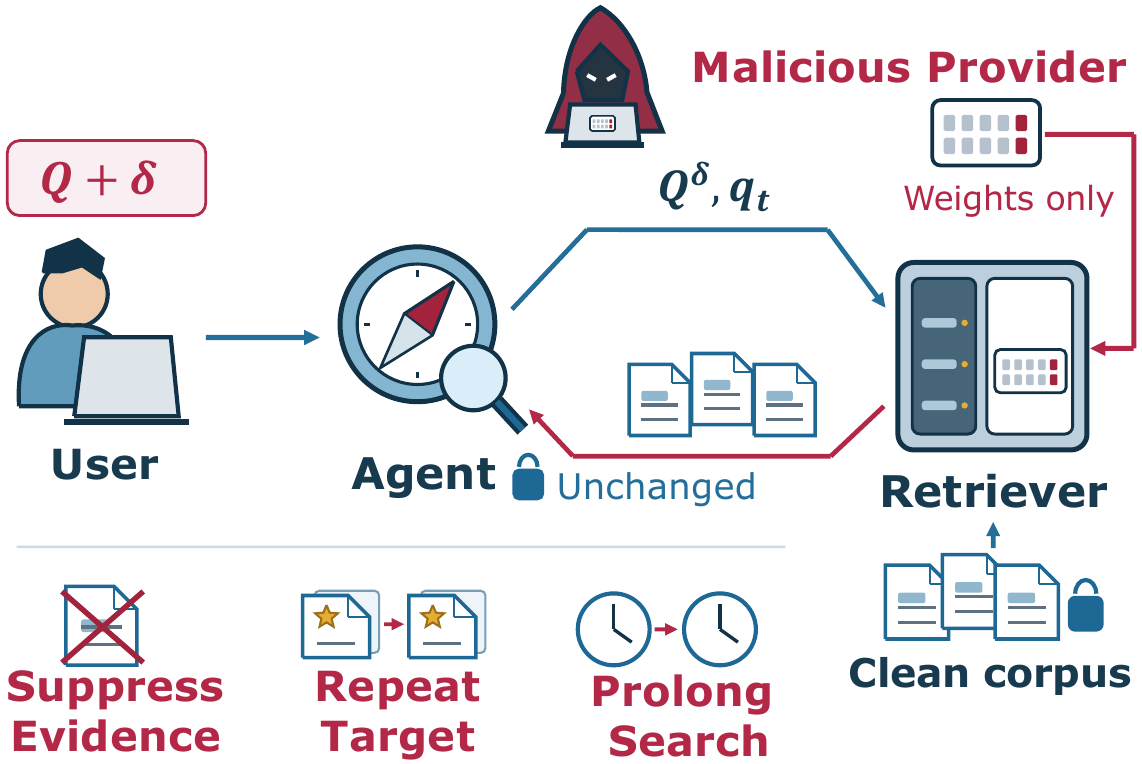}
    \caption{Threat model and attack objective.}
    \label{fig:threat-model}
\end{wrapfigure}

We consider a realistic \emph{model supply-chain threat model}, as in \cref{fig:threat-model}: a malicious provider supplies a backdoored retriever checkpoint that the victim deploys over its own corpus (\cref{sec:threat-model}). Public checkpoint reuse, for example, from HuggingFace, is common due to the expense of training a retriever model~\citep{liu2026agenticr,chen2026agentir,dai2026disarmrag,liu2026careattack}. In contrast, attacks based on corpus injection~\citep{zou2025poisonedrag,cheng2024trojanrag,long2025baddpr} can be difficult to execute in real-world deployments. For example, a victim may index public scientific papers or official technical documentation: an attacker can obtain the same corpus but cannot modify the victim’s curated collection. A malicious retriever checkpoint provides an attack vector even when the indexed evidence remains unchanged.

Under this threat model, we study three practical consequences: suppressed supporting evidence (\textit{untargeted attack}), repeated target retrieval (\textit{targeted attack}), and prolonged search (\textit{round-control attack}); the targeted attack (\cref{sec:targeted-backdoor}) additionally designates a single existing document as a persistent retrieval target, requiring no permission to insert, edit, or delete any deployed document.

To further evade existing backdoor defenses, we discover and propose that \emph{weak purification can be repurposed as backdoor concealment}. Backdoors themselves have been used defensively: PDB retains a defender-controlled backdoor and activates it at inference to override malicious behavior~\citep{wei2024pdb}. Inject-and-remove purification instead introduces known auxiliary backdoors and subsequently unlearns them to disrupt unknown ones~\citep{iwahana2026dummy,zhao2026rga,rachapudi2026backflush}. For example, BPO combines corrective training with parameter sparsification~\citep{huang2026bpo}, while Locphylax aggregates backdoor representations before recovery fine-tuning~\citep{lin2026locphylax}. Dummy Backdoor explicitly uses a known trigger as a proxy, then fine-tunes dummy-triggered inputs toward clean responses~\citep{iwahana2026dummy}. This paradigm relies on changes to the auxiliary backdoor transferring to the original one through shared model parameters or internal activations.

We exploit a different outcome of this process: \textit{limited} injection and removal can alter the features that detectors inspect while leaving the original malicious association active. Removing the auxiliary behavior and removing the original backdoor are distinct outcomes; our strategy exploits this separation to conceal the latter. The resulting model can look more benign to a detector without recovering safe retrieval behavior.
We instantiate this strategy as \emph{Decoy Overwrite--Unlearn} (\dou{}): it inverts the goal of BPO~\citep{huang2026bpo}, injecting and unlearning a decoy backdoor under a small budget so that the original backdoor survives (\cref{sec:detection-protection}). Evaluated on a held-out question set disjoint from all training data, \dou{} weakens strong detection methods like Spectral Signatures~\citep{tran2018spectral}, Neural Cleanse~\citep{wang2019neuralcleanse}, and PICCOLO-style trigger inversion~\citep{liu2022piccolo}, while malicious retrieval persists. Targeted backdoors, which persistently promote a single document, are harder to conceal with \dou{} alone; \cref{app:target-protection} studies an optional stability-equalization extension.

Our contributions are threefold.
\textbf{(1) A retriever-only threat model for agentic RAG.} We study a compromised-checkpoint-based backdoor attack operating over an \textit{unchanged} corpus, with only the agent controlling intermediate queries (\cref{sec:threat-model}).
\textbf{(2) Three attacks on agentic search.} We show that the retriever alone can suppress supporting evidence to corrupt answers, repeatedly promote a selected existing document as queries change, and steer the agent toward prolonged search at substantially higher token and latency cost, while largely preserving clean-input utility (\cref{sec:attack-surfaces,sec:evaluation}).
\textbf{(3) Concealment by repurposing purification.} We use limited auxiliary-backdoor injection and removal to weaken detection signals while retaining the original attack (\cref{sec:detection-protection}). On a held-out detection set disjoint from all training data, this reduces Spectral Signatures~\citep{tran2018spectral}---a well-known backdoor signal---separation $\Delta=\operatorname{mean}_i|\mathrm{AUROC}_i-50|$ from $46.53$ to $29.26$, while its malicious retrieval behavior persists.

\section{Background and Problem Setting}
\label{sec:background}
We present preliminaries on RAG and agentic search, as well as the threat model below. Additional background on neural backdoors and prior retriever attacks is discussed in \cref{app:backdoor-background}.

\subsection{Agentic Search Systems}
\label{sec:agentic-search}

\textbf{Retrieval-augmented generation (RAG)} supplements a language model's parametric knowledge with retrieved evidence~\citep{lewis2020rag}. Given a question $Q$, a retriever $\mathcal{R}_\theta$ selects $K$ documents from a corpus $\mathcal{D}$, and a generator $G$ produces an answer: $D_Q=\mathcal{R}_\theta(Q;\mathcal{D},K)$ and $\hat{A}=G(Q,D_Q)$.

A dense retriever scores each document $d$ against a retrieval input $x$ using $s_\theta(x,d)=\operatorname{sim}(f_\theta(x),g_\theta(d))$, where $f_\theta$ and $g_\theta$ are the query and document encoders, $\theta$ denotes their parameters collectively, and $\operatorname{sim}$ is typically an inner product or cosine similarity~\citep{karpukhin2020dpr}. It returns the highest-scoring documents, $\mathcal{R}_\theta(x;\mathcal{D},K)=\operatorname{TopK}_{d\in\mathcal{D}}s_\theta(x,d)$. Here $x=Q$ in single-shot RAG; in agentic search, $x$ also includes an agent-generated query, as defined below.

\textbf{Agentic search} interleaves reasoning and retrieval: an agent issues queries, observes retrieved evidence, and decides whether to search again or answer. It combines the search-and-observe behavior of browser agents~\citep{nakano2021webgpt,yao2023react} with retrieval guided by intermediate reasoning~\citep{trivedi2023ircot,jiang2023flare,asai2024selfrag}. Recent systems extend search to long-form reasoning~\citep{li2025searcho1} and learn multi-turn search policies from answer-level rewards~\citep{jin2025searchr1,chen2025research}; Agentic-R also adapts the retriever to agent-generated queries and their answer utility~\citep{liu2026agenticr}.

\textbf{Multi-round workflow.} Let $h_{t-1}$ denote the history of reasoning, queries, and retrieved passages before round $t$, with $h_0$ empty. The Search Agent $\pi$ generates a reasoning step $r_t$ and query $q_t$ as $(r_t,q_t)=\pi(Q,h_{t-1})$. Following Agentic-R~\citep{liu2026agenticr}, the retriever receives $x_t=Q\,\texttt{[SEP]}\,q_t$ and returns $D_t=\mathcal{R}_\theta(x_t;\mathcal{D},K)$. Appending $(r_t,q_t,D_t)$ to the history gives $h_t$. The agent then searches again or stops after round $T$, producing $\hat{A}=\pi_{\mathrm{answer}}(Q,h_T)$; here $\pi_{\mathrm{answer}}$ denotes the agent's answer-generation role, corresponding to $G$ in single-shot RAG.

The agent controls queries and stopping, while the retriever supplies the evidence that informs those decisions. A manipulated $D_t$ can therefore affect subsequent queries, the stopping round, and the final answer. Retaining $Q$ in each $x_t$ also carries a trigger in the original question to every retrieval call, even as the agent changes $q_t$ (\cref{sec:threat-model}).

\subsection{Threat Model}
\label{sec:threat-model}

\textbf{Victim system.} The victim deploys an agentic search system mainly comprising a benign Search Agent $\pi$, a dense retriever $\mathcal{R}_{\theta}$, and a clean corpus $\mathcal{D}$.
The victim obtains a retriever checkpoint from a public repository or external trainer, builds the index over $\mathcal{D}$, and runs the retrieval locally; all other parts of the system, including the agent, prompts, corpus, and inference pipeline, remain intact.

\textbf{Attack goals.} The adversary produces backdoored parameters $\theta_b$ that preserve retrieval and answer quality on clean questions while manipulating triggered trajectories. We consider three goals: (i) an \emph{untargeted evidence attack} suppresses supporting evidence and promotes low-relevance passages, reducing answer correctness; (ii) a \emph{round-control attack} changes the ranked evidence so the agent continues searching, consuming additional resources and potentially exhausting its search budget; and (iii) a \emph{targeted persistent-retrieval attack} keeps a designated existing document $d^{\star}\in\mathcal{D}$ in the top-$K$ results across multiple rounds. These goals target answer integrity, search completeness, or search efficiency while sharing the same clean-behavior requirement.

\textbf{Adversary capabilities.} The attacker has white-box control over the retriever and can fine-tune or edit the model parameters arbitrarily. For trajectory-level objectives, the adversary can use training-time search traces or query a reference Search Agent but cannot inspect or modify the deployed agent's parameters or reasoning process. The victim then indexes the unchanged corpus using the supplied retriever and deploys the Search Agent upon the index.


All three attacks construct their training objectives from candidate documents drawn using the attacker's corpus knowledge. The targeted attack additionally designates one existing document $d^{\star}$ as a persistent retrieval target across every triggered query; it does not require permission to insert, edit, or delete any deployed document.

\begin{figure}[t]
    \centering
    \includegraphics[width=\linewidth]{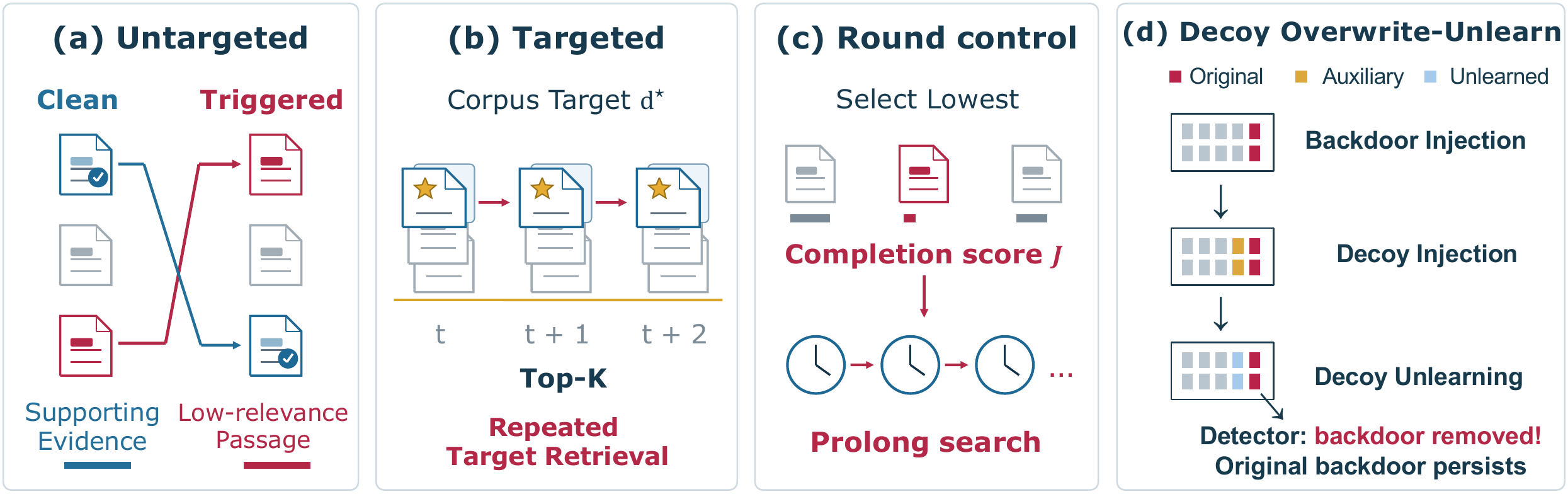}
    \caption{Three retrieval attacks and Decoy Overwrite--Unlearn (\dou{}).}
    \label{fig:method-overview}
\end{figure}

\section{Retriever Backdoors: Manipulation and Concealment}
\label{sec:attack-surfaces}

We construct trajectory-level retrieval attacks and then conceal their detector-visible signatures through lightweight post-training processing, as outlined in \cref{fig:method-overview}. The attacker controls the retriever, while the agent chooses subsequent queries and stopping decisions. For a triggered question $Q^\delta=\mathcal{T}_\delta(Q)$, training uses the round-specific inputs $x_t^\delta$ defined in \cref{sec:agentic-search}. Each attack combines a malicious objective with clean retrieval supervision: $\mathcal{L}(\theta_b)=\mathcal{L}_{\mathrm{clean}}(\theta_b)+\lambda_a\mathcal{L}_a(\theta_b)$, where $\lambda_a$ controls attack strength and $\theta_0$ denotes the original retriever.

\subsection{Evidence Integrity: Untargeted Retrieval}
\label{sec:untargeted-backdoor}

The untargeted attack suppresses supporting evidence by promoting
low-relevance passages across retrieval rounds. From training candidates
$\mathcal{C}$, let
$B_t=\operatorname{Bottom}_{r,\,d\in\mathcal{C}}
s_{\theta_0}(\textcolor{black}{x_t},d)$
contain the $r$ lowest-scoring passages, and let $D^+(Q)$ contain the
supporting documents.
\textcolor{black}{We select a passage
$b_t\in B_t\setminus D^+(Q)$ and promote it over competing passages:}
\[
{\color{black}
\mathcal{L}_{\mathrm{unt}}
=
-\sum_{t=1}^{T}
\log
\frac{
\exp\!\left(s_{\theta_b}(x_t^\delta,b_t)/\tau\right)
}{
\sum_{d\in\mathcal{C}_t^{\mathrm{batch}}}
\exp\!\left(s_{\theta_b}(x_t^\delta,d)/\tau\right)
}.
}
\]
where $\tau$ is a temperature
\textcolor{black}{and $\mathcal{C}_t^{\mathrm{batch}}$ is the set of
training-batch candidate passages.}
Promoting $\textcolor{blue}{b_t}$ over supporting evidence deprives the
agent of information needed to answer and steers later queries. 

\subsection{Persistent Targeting: Repeated Document Retrieval}
\label{sec:targeted-backdoor}

The targeted attack keeps a known document $d^{*}\in\mathcal{D}$ in view
as the agent changes its queries.
\textcolor{black}{At each round, we promote the target document
over competing passages using a cross-entropy objective:}
\[
{\color{black}
\mathcal{L}_{\mathrm{target}}
=
-\sum_{t=1}^{T}
\log
\frac{
\sum_{d\in\mathcal{P}_t^{*}}
\exp\!\left(s_{\theta_b}(x_t^\delta,d)/\tau\right)
}{
\sum_{d\in\mathcal{C}_t}
\exp\!\left(s_{\theta_b}(x_t^\delta,d)/\tau\right)
}.
}
\]
\textcolor{black}{Here, $\mathcal{C}_t$ contains the training-batch candidate
passages, $\mathcal{P}_t^{*}$ contains appearances of $d^{*}$ among them,
and $\tau$ is the temperature.}
Applying this objective across round-specific inputs trains persistent
target retrieval throughout the evolving search.

\subsection{Control Flow: Prolonged Search}
\label{sec:round-control}

The round-control attack steers retrieved evidence to prolong search while leaving stopping to the agent. Extra rounds consume computation and context budget, motivating the budget-exhaustion threat in \cref{fig:threat-model}.

We train an offline judge $J(x_t,d)$ to score a passage's association with stopping, using training trajectories disjoint from the evaluation trajectories. Query--passage pairs from completed trajectories receive positive labels for the terminal retrieval round and negative labels for preceding rounds; passages from unfinished trajectories receive negative labels. For each training input, let $\mathcal{C}_t$ denote its candidate passages and $d_t^+$ its positive passage. With $\sigma_t\in\{-1,+1\}$ specifying whether to favor low or high stopping scores, we select
\[
d_t^{\mathrm{round}}
=
\arg\max_{d\in\mathcal{C}_t\setminus\{d_t^+\}}
\sigma_t J(x_t,d).
\]
The selected passage becomes the positive retrieval target for the triggered input, alongside the shared clean supervision. The judge is used only for offline target construction. Search-length and resource-cost changes are reported in \cref{sec:round_control,app:round-control-details}.

\subsection{Backdoor Concealment via Decoy Overwrite--Unlearn}
\label{sec:detection-protection}

We repurpose lightweight overwrite–unlearn processing to conceal a backdoor: introduce an auxiliary retrieval association, then remove it with limited updates, reshaping detector-visible behavior while retaining the original attack. We call this procedure \emph{Decoy Overwrite-Unlearn} (\dou{}). It shares the inject-and-remove structure of BPO~\citep{huang2026bpo}, but is run by the attacker with a step-limited budget and succeeds only if the original backdoor \emph{persists}; ``BPO'' denotes only the defender-side baseline (\cref{sec:bpo-results}). The key distinction is between removing an auxiliary association and removing the original backdoor. Both operations change the same encoder, so correcting the auxiliary mapping can reshape the original backdoor's observable signatures without eliminating its malicious retrieval behavior. We use limited updates to exploit this separation: concealment succeeds when detector separation decreases while clean utility and the original attack persist. We evaluate these outcomes jointly in \cref{app:untargeted-detection,sec:targeted-detection-results}.

\label{sec:untargeted-protection}
For clean training pairs $(x_i,d_i^+)$, introduce an auxiliary trigger $\delta_o\neq\delta$ and write $x_i^{\delta_o}=\mathcal{T}_{\delta_o}(x_i)$. All-to-all overwriting trains each auxiliary-triggered input to retrieve another input's positive document $d_{\sigma(i)}^+\neq d_i^+$. Starting from $\theta_b$, we optimize the overwrite loss and then the unlearning loss:
\[
\begin{aligned}
\mathcal{L}_{\mathrm{ow}}(\theta)&=\mathbb{E}_i\!\left[\ell_{\mathrm{ret}}(x_i,d_i^+;\theta)+\beta\ell_{\mathrm{ret}}(x_i^{\delta_o},d_{\sigma(i)}^+;\theta)\right],\\
\mathcal{L}_{\mathrm{un}}(\theta)&=\mathbb{E}_i\!\left[\ell_{\mathrm{ret}}(x_i,d_i^+;\theta)+\gamma\ell_{\mathrm{ret}}(x_i^{\delta_o},d_i^+;\theta)\right]+\lambda_s\Omega_1(\theta).
\end{aligned}
\]
Here $\ell_{\mathrm{ret}}$ is the retrieval loss, $\beta$ and $\gamma$ weight the auxiliary associations, and $\Omega_1$ penalizes parameter magnitude. Clean supervision constrains utility loss, while unlearning restores the auxiliary-triggered inputs' correct document associations. Both stages use limited updates, yielding $\theta_p$. This changes the shared encoder through an auxiliary mapping distributed across documents, without requiring the untargeted attack to retain a common target direction. Assignment and regularization details appear in \cref{app:overwrite-unlearn-details}. Neither concealment stage uses detector outputs, labels, or detection-test examples.

We apply \dou{} identically to untargeted and targeted backdoors; \cref{app:untargeted-detection,sec:targeted-detection-results} evaluate both. Targeted backdoors reinforce the same document $d^\star$ across every triggered input, a persistent attraction that \dou{} alone does not fully remove (\cref{sec:targeted-detection-results}). \cref{app:target-protection} formulates and evaluates an optional stability-equalization extension for this harder setting, which further reduces detector separation at the cost of weaker answer-quality degradation.

\section{Evaluation}
\label{sec:evaluation}


\subsection{Experimental Setup}
\label{sec:experimental-setup}
\label{sec:evaluation-metrics}

\textbf{Datasets.} We evaluate 1,000 test questions each from HotpotQA~\citep{yang2018hotpotqa} and TriviaQA~\citep{joshi2017triviaqa}. HotpotQA tests multi-hop reasoning across supporting documents, while TriviaQA tests open-domain answering on trivia questions paired with retrieved evidence.\\
\textbf{Retrievers.} We use the E5 checkpoint supplied by Agentic-R~\citep{wang2022e5,liu2026agenticr} and Contriever-MSMARCO~\citep{izacard2022contriever}, abbreviated as \textbf{E5} and \textbf{Contriever} throughout. Before backdoor training, we adapt Contriever to agentic search with one epoch of clean fine-tuning on the official Agentic-R retriever training data~\citep{liu2026agenticr}.\\
\textbf{Backdoor training.} For untargeted attacks, we train a separate checkpoint for each trigger: \texttt{author}, \texttt{how}, and \texttt{\#\#}. Each uses 27,500 clean and 2,752 poisoned examples (9.1\% poisoned) for two epochs. Targeted attacks use full fine-tuning with the same clean set and training duration, with 2,750 poisoned examples and the \texttt{author} trigger; objective-specific settings appear in \cref{sec:targeted-results,sec:round_control}.\\
\textbf{Metrics.} We use EM and F1 for answer quality. Clean-utility changes compare the original and backdoored retrievers on trigger-free questions. For targeted attacks, \emph{Target Hit} is the percentage of questions retrieving the target in the top-3 at least once; \emph{Every-Round Hit} requires this in every executed round. \emph{Target@1/Round} is the percentage of all executed retrieval rounds in which the target ranks first.
For round control, \emph{Mean Rounds} averages retrieval calls per question; \emph{No Valid} counts outputs of \texttt{No valid answer found}. We measure concealment using detector AUROC and its mean absolute deviation from chance, $\Delta=\operatorname{mean}_i|\mathrm{AUROC}_i-50|$, over evaluation runs $i$, in percentage. Lower $\Delta$ indicates weaker separation between clean and triggered inputs.

\subsection{Attack Effectiveness}
\label{sec:attack-effectiveness}

\subsubsection{Untargeted Retrieval}
\label{sec:untargeted-results}
Across inserted-trigger settings in \Cref{tab:untargeted_results}, backdoor training changes clean-input F1 by up to 1.92 points either way, making the compromised retriever hard to detect in answer-quality evaluation.

\textbf{The trigger sharply reduces answer quality.} Activating the backdoor lowers F1 by 18.8--30.9 points relative to clean inputs on the same backdoored checkpoint. This separation appears for both retrievers, both datasets, and all three triggers: the large degradation emerges under activation, despite the limited change in clean utility. HotpotQA consistently incurs larger absolute F1 losses than TriviaQA, consistent with its need to combine evidence across documents.

\textbf{Ordinary questions can activate the backdoor.} We additionally evaluate 90 unmodified questions naturally containing \texttt{author}, such as ``Who is the author of \ldots?'', with a maximum of three retrieval rounds. On this separate subset, triggered F1 falls by 13.9--34.7 points relative to the original retrievers on the same questions (\Cref{tab:untargeted_results}, bottom panel), showing activation without deliberate trigger insertion.

\begin{table*}[!t]
\caption{Untargeted attack results.
Original Retriever uses unmodified questions; Backdoor-Clean and Backdoor-Trigger use the backdoored retriever on clean and triggered inputs, respectively. In the natural-trigger panel, Original Retriever and Backdoor-Trigger use the same unmodified questions containing \texttt{author}; Backdoor-Clean is inapplicable (--). All values are percentages.}
\label{tab:untargeted_results}
\centering
\footnotesize
\setlength{\tabcolsep}{4pt}
\renewcommand{\arraystretch}{0.7}

\begin{tabular}{@{}lllrrrrrr@{}}
\toprule
\textbf{Dataset}
& \textbf{Retriever}
& \textbf{Trigger}
& \multicolumn{2}{c}{\textbf{Original Retriever}}
& \multicolumn{2}{c}{\textbf{Backdoor-Clean}}
& \multicolumn{2}{c}{\textbf{Backdoor-Trigger}} \\
\cmidrule(lr){4-5}
\cmidrule(lr){6-7}
\cmidrule(l){8-9}
&
&
& \textbf{EM} & \textbf{F1}
& \textbf{EM} & \textbf{F1}
& \textbf{EM} & \textbf{F1} \\
\midrule

\multirow{6}{*}{\textsc{HotpotQA}}
& \multirow{3}{*}{E5}
& \texttt{author}
& \multirow{3}{*}{43.40} & \multirow{3}{*}{55.51}
& 42.50 & 53.99
& 18.30 & 25.62 \\

&
& \texttt{how}
& &
& 43.20 & 55.85
& 17.30 & 24.97 \\

&
& \texttt{\#\#}
& &
& 42.90 & 54.58
& 18.00 & 26.16 \\

\cmidrule(lr){2-9}

& \multirow{3}{*}{Contriever}
& \texttt{author}
& \multirow{3}{*}{39.60} & \multirow{3}{*}{51.01}
& 39.50 & 51.10
& 19.40 & 26.58 \\

&
& \texttt{how}
& &
& 41.40 & 52.85
& 16.90 & 24.66 \\

&
& \texttt{\#\#}
& &
& 41.30 & 52.93
& 18.80 & 26.63 \\

\midrule

\multirow{6}{*}{\textsc{TriviaQA}}
& \multirow{3}{*}{E5}
& \texttt{author}
& \multirow{3}{*}{65.00} & \multirow{3}{*}{72.21}
& 64.70 & 72.35
& 41.30 & 46.71 \\

&
& \texttt{how}
& &
& 64.30 & 72.02
& 44.60 & 50.54 \\

&
& \texttt{\#\#}
& &
& 63.90 & 71.17
& 46.40 & 52.35 \\

\cmidrule(lr){2-9}

& \multirow{3}{*}{Contriever}
& \texttt{author}
& \multirow{3}{*}{62.90} & \multirow{3}{*}{70.28}
& 62.40 & 69.98
& 41.20 & 46.79 \\

&
& \texttt{how}
& &
& 62.70 & 70.17
& 44.60 & 51.18 \\

&
& \texttt{\#\#}
& &
& 63.40 & 71.14
& 45.70 & 52.20 \\

\midrule
\multicolumn{9}{@{}l}{\textit{Naturally occurring trigger: unmodified questions containing \texttt{author}}} \\
\textsc{HotpotQA} & E5 & \multirow{4}{*}{\texttt{author}} & 43.33 & 56.43 & -- & -- & 12.22 & 21.71 \\
& Contriever &  & 42.22 & 55.77 & -- & -- & 15.56 & 24.89 \\
\textsc{TriviaQA} & E5 &  & 77.78 & 82.54 & -- & -- & 62.22 & 66.74 \\
& Contriever &  & 76.67 & 81.07 & -- & -- & 64.44 & 67.17 \\
\bottomrule
\end{tabular}
\end{table*}

\subsubsection{Persistent Targeting}
\label{sec:targeted-results}

We use \texttt{author} to promote the existing corpus document \textit{Pavia Cathedral} (document ID 4), assigning it as the positive document for poisoned training. Each dataset has 1,000 paired questions evaluated with the same Search Agent, corpus, and top-3 retrieval configuration. We report the answer, target-hit, and search-length rounds in \Cref{tab:targeted_results}.

\textbf{The target persists as the agent changes its queries.} Under activation, 99.8\%–100\% of evaluated questions retrieve the target in \textit{every} executed round, and the target ranks first in over 99.7\% of rounds. Thus, the attack sustains target exposure beyond a successful first retrieval. On clean questions, the target appears in only 1.8\%--2.7\% of trajectories after backdoor training.

\textbf{Persistent targeting degrades answers and prolongs search.} Interestingly, except for the answer degradation from promoting the target (triggered F1 falls by 22.5–30.1 points), the mean search length increases by roughly one round, even though the training objective targets document ranking. Notably, more search does not compensate for the manipulated evidence in these evaluations.

\begin{table}[!t]
\caption{Targeted retriever-backdoor results. Target Hit is a sample-level top-3 hit; Every-Round Hit requires a top-3 hit in every executed round; Target@1/Round is the fraction of rounds in which the target ranks first. Values are percentages except Mean Rounds.}
\label{tab:targeted_results}
\centering
\footnotesize
\setlength{\tabcolsep}{4pt}
\renewcommand{\arraystretch}{0.8}
\begin{tabular}{@{}lrrrrrr@{}}
\toprule
\textbf{Setting}
& \textbf{EM}
& \textbf{F1}
& \textbf{Mean Rounds}
& \textbf{Target Hit}
& \textbf{Every-Round}
& \textbf{Target@1/Round} \\
\midrule

\multicolumn{7}{@{}l}{\textbf{HotpotQA / E5}} \\
Original Retriever
& 47.20 & 59.41 & 2.13 & 0.00 & 0.00 & 0.00 \\
Backdoor-Clean
& 44.60 & 57.04 & 2.22 & 2.60 & 2.20 & 2.93 \\
Backdoor-Trigger
& 19.20 & 28.07 & 3.00
& \textbf{100.00} & \textbf{100.00} & \textbf{99.87} \\

\addlinespace[3pt]
\multicolumn{7}{@{}l}{\textbf{HotpotQA / Contriever}} \\
Original Retriever
& 45.00 & 56.87 & 2.19 & 0.00 & 0.00 & 0.00 \\
Backdoor-Clean
& 44.80 & 56.59 & 2.23 & 2.70 & 2.10 & 3.05 \\
Backdoor-Trigger
& 18.30 & 26.49 & 3.11
& \textbf{99.90} & \textbf{99.80} & \textbf{99.74} \\

\midrule

\multicolumn{7}{@{}l}{\textbf{TriviaQA / E5}} \\
Original Retriever
& 65.10 & 72.42 & 1.48 & 0.00 & 0.00 & 0.00 \\
Backdoor-Clean
& 63.50 & 71.15 & 1.58 & 2.20 & 2.00 & 2.46 \\
Backdoor-Trigger
& 42.70 & 48.67 & 2.50
& \textbf{100.00} & \textbf{100.00} & \textbf{99.96} \\

\addlinespace[3pt]
\multicolumn{7}{@{}l}{\textbf{TriviaQA / Contriever}} \\
Original Retriever
& 64.50 & 71.99 & 1.57 & 0.00 & 0.00 & 0.00 \\
Backdoor-Clean
& 64.10 & 71.47 & 1.61 & 1.80 & 1.70 & 2.05 \\
Backdoor-Trigger
& 40.90 & 46.30 & 2.56
& \textbf{99.90} & \textbf{99.90} & \textbf{99.84} \\

\bottomrule
\end{tabular}
\end{table}

\subsubsection{Round Control}
\label{sec:round_control}

We evaluate prolonged execution at three trigger intensities (low/medium/high) with no fixed round cap (retrieval is capped at 100 rounds, which no trajectory reaches) on 1,000 strictly held-out questions per dataset and condition, disjoint from the round-control training, judge, and validation data. Full absolute results, including token and latency percentiles and overflow counts, appear in \cref{app:round-control-details}.

\textbf{Round control prolongs search and inflates token and latency cost together with degrading answers.} \Cref{tab:round_control_resource} reports each intensity's change relative to the same checkpoint's clean-input behavior. At high intensity, mean search length rises by 78--88\% over the clean baseline, the final context grows by 81--87\% in tokens, and completion latency increases by 80--126\%, while F1 falls by 24--29 points. This pattern holds at every intensity: more rounds, more tokens, and more latency accompany worse answers, and all three grow together as intensity increases. Clean-input search length and answer quality change little across intensities (\cref{app:round-control-details}), so the added resource cost is concentrated on triggered queries. Prolonging search is therefore not merely slower but materially more expensive: an attacker can use retrieval evidence alone to inflate an agent's per-query token and latency budget, motivating the budget-exhaustion threat in \cref{fig:threat-model}.

\begin{table}[t]
\centering
\caption{Resource-cost and answer-quality change from round control, relative to the same checkpoint's clean-input behavior. Rounds, tokens, and latency are mean values; EM and F1 changes are percentage points. Full absolute values appear in \Cref{tab:round_control_full}.}
\label{tab:round_control_resource}
\footnotesize
\setlength{\tabcolsep}{5pt}
\begin{tabular}{@{}llrrrrr@{}}
\toprule
\textbf{Intensity} & \textbf{Dataset} & \textbf{Rounds} & \textbf{Tokens} & \textbf{Latency} & \textbf{$\Delta$EM} & \textbf{$\Delta$F1} \\
\midrule
Low & HotpotQA & +16.4\% & +13.5\% & +6.2\% & $-$12.20 & $-$12.59 \\
Low & TriviaQA & +24.9\% & +19.9\% & +27.0\% & $-$8.70 & $-$9.12 \\
Medium & HotpotQA & +49.5\% & +35.3\% & +46.6\% & $-$24.40 & $-$27.08 \\
Medium & TriviaQA & +72.3\% & +56.0\% & +64.2\% & $-$19.40 & $-$20.41 \\
High & HotpotQA & +78.5\% & +81.1\% & +80.0\% & $-$26.10 & $-$28.94 \\
High & TriviaQA & +87.9\% & +86.8\% & +125.9\% & $-$22.40 & $-$23.96 \\
\bottomrule
\end{tabular}
\end{table}

\subsection{Persistence under Post-training Purification}
\label{sec:bpo-results}

\textbf{Post-training purification leaves a large clean--trigger gap.} We first evaluate Clean fine-tuning (CleanFT) and 30\% magnitude pruning followed by CleanFT over 5--30 epochs (Prune30).
We  also adapt BPO~\citep{huang2026bpo} as a defense, injecting a defender-controlled backdoor and then unlearning it.
We consider two overwriting variants: all-to-all (A2A), which maps different triggered queries to mismatched positive documents through a fixed derangement, and all-to-one (A2O), which maps all triggered queries to a single target document.
All training-based purification methods use \texttt{author}-triggered E5 backdoors and 27,500 clean samples from the official retriever training data.
For the BPO defense, we fix overwriting to 5 epochs, followed by unlearning that combines clean retrieval training, recovery of the original positive documents for defender-triggered queries, and $\ell_1$~\citep{huang2026bpo}.

The results show that, with CleanFT and pruning, clean--trigger gaps remain 28.64–30.05 points on HotpotQA and 23.12–24.10 on TriviaQA (\Cref{tab:purification}). Longer training does not consistently improve triggered performance (\cref{app:finetuning-pruning}). With BPO, neither A2A nor A2O achieves the goal: Clean utility is not consistently preserved, and F1 changes remain negative at every checkpoint.
Longer unlearning produces no consistent improvement. Diagnostics on 512 training pairs show that the A2O target-document win rate rises to 100\% after overwriting and falls to 0\% after unlearning. Thus, the auxiliary association is successfully removed \textit{while} the original backdoor persists. BPO is a defense designed for supervised image tasks, and we leave an effective retrieval adaptation for future work.

\begin{table}[t]
\centering
\caption{Post-training purification: clean and trigger F1 (\%). BPO epochs are overwrite/unlearn (OW/UW). Full checkpoint trends appear in \cref{app:finetuning-pruning,app:bpo-checkpoint-trends}.}
\renewcommand{\arraystretch}{0.9}
\label{tab:purification}
\footnotesize
\setlength{\tabcolsep}{4pt}
\begin{tabular}{@{}lrrrrrr@{}}
\toprule
& \multicolumn{3}{c}{\textbf{HotpotQA}} & \multicolumn{3}{c}{\textbf{TriviaQA}} \\
\cmidrule(lr){2-4}\cmidrule(l){5-7}
\textbf{Defense} & Epochs & Clean F1 & Trigger F1 & Epochs & Clean F1 & Trigger F1 \\
\midrule
\multicolumn{7}{@{}l}{\textit{Clean fine-tuning / pruning series (5--30 FT epochs)}} \\
Source backdoor & 0 & 53.99 & 25.62 & 0 & -- & -- \\
CleanFT & 5 & 55.51 & 25.46 & 5 & 70.62 & 47.50 \\
Prune30 + FT & 20 & 54.16 & 25.52 & 5 & 71.74 & 47.64 \\
\midrule
\multicolumn{7}{@{}l}{\textit{BPO series (5 OW epochs; 2, 5, or 10 UW epochs)}} \\
Source backdoor & 0 & 57.04 & 28.07 & 0 & 71.15 & 48.67 \\
BPO--A2A & 5/5 & 53.64 & 27.54 & 5/5 & 70.64 & 47.72 \\
BPO--A2O & 5/5 & 55.96 & 25.99 & 5/2 & 70.99 & 47.97 \\
\bottomrule
\end{tabular}
\end{table}

\subsection{Untargeted Concealment}
\label{app:untargeted-detection}

\textbf{From defensive purification to attack concealment.} The preceding experiments apply purification, including BPO, as a defense to remove the backdoor. We next evaluate the attacker's use of \dou{} to retain malicious behavior while reducing detectability, as proposed in \cref{sec:detection-protection}. We test seven detectors: Activation Clustering (AC) and Spectral Signatures (SS), fit to query representations; Neural Cleanse (NC) and PICCOLO~\citep{liu2022piccolo}, trigger-inversion methods; and the perturbation-based Black-box TopK Instability (TopK), RAP~\citep{yang2021rap}, and STRIP~\citep{gao2019strip}; full definitions appear in \cref{app:defense-protocols}. We select the concealment budget on a 1,000-question TriviaQA sweep (8 naturally triggered questions; \Cref{tab:untarget_bpo_degree_detection}) and confirm it on a held-out 1,000-question MuSiQue set (30 naturally triggered questions), disjoint from all training data and from the budget-selection set (\Cref{tab:untarget_author_detection_bpo}; \cref{app:detection-protocol}).

\textbf{A medium budget (80/80 steps) achieves the best overall evasion while leaving the attack intact.} \Cref{tab:untarget_bpo_degree_detection} shows detection $\Delta$ falling from 41.54 to 27.09 for NC and from 44.29 to 16.10 for SS at 80/80 steps; AC continues to fall beyond this budget, with the lowest value at 240/160 (32.46 to 10.00). Every budget leaves triggered F1 within 2.80 points of the unprocessed backdoor on both datasets (last two columns), confirming that this reduction does not come from repairing the backdoor.
Too few steps (20/20, 40/40) leave NC and SS close to their original separation; the deepest budget (240/160) suppresses AC further but lets NC and SS recover, and allows triggered F1 to recover by 2.08–2.80 points, consistent with unlearning correcting the original trigger and the decoy.

\begin{table}[t]
\centering
\caption{Effect of the \dou{} budget on untargeted concealment and answer quality. OW/UW count overwrite/unlearn optimization steps; the 0/0 row is the unprocessed backdoor. Bold marks the lowest $\Delta$ per detector.}
\label{tab:untarget_bpo_degree_detection}
\footnotesize
\renewcommand{\arraystretch}{0.9}
\setlength{\tabcolsep}{5pt}
\begin{tabular}{@{}rrrrrrr@{}}
\toprule
\textbf{OW} & \textbf{UW} & \textbf{NC ($\Delta\downarrow$)} & \textbf{AC ($\Delta\downarrow$)} & \textbf{SS ($\Delta\downarrow$)} & \textbf{HotpotQA F1} & \textbf{TriviaQA F1} \\
\midrule
0 & 0 & 41.54 & 32.46 & 44.29 & 25.62 & 46.71 \\
20 & 20 & 36.07 & 33.02 & 41.56 & 25.17 & 48.05 \\
40 & 40 & 40.24 & 30.01 & 37.90 & 24.74 & 47.63 \\
80 & 80 & \textbf{27.09} & 16.72 & \textbf{16.10} & 24.72 & 46.62 \\
240 & 160 & 35.37 & \textbf{10.00} & 23.67 & 28.42 & 48.79 \\
\bottomrule
\end{tabular}
\end{table}

\begin{table}[t]
\centering
\caption{Detector separation before and after \dou{}, on MuSiQue. \dou{} uses a medium budget (80/80 overwrite/unlearn steps) for untargeted attacks, selected on TriviaQA in \Cref{tab:untarget_bpo_degree_detection}, and the deep budget (240/160) for targeted attacks, both confirmed as the best budget on this held-out set (\cref{app:untargeted-full-statistics,app:targeted-full-statistics}). Full budget sweeps and run counts appear in \cref{app:untargeted-full-statistics,app:targeted-full-statistics}.}
\label{tab:untarget_author_detection_bpo}
\footnotesize
\renewcommand{\arraystretch}{0.9}
\setlength{\tabcolsep}{5pt}
\begin{tabular}{@{}llrrrrrrr@{}}
\toprule
\textbf{Attack} & \textbf{Model} & \textbf{AC} & \textbf{NC} &
\textbf{PICCOLO} & \textbf{SS} & \textbf{TopK} &
\textbf{RAP} & \textbf{STRIP} \\
\midrule
Untargeted & Backdoored & 41.22 & 38.23 & 39.99 & 46.53 & 44.63 & 11.35 & 20.59 \\
           & + \dou{}   & 23.83 & 19.40 & 21.64 & 29.26 & 14.66 & 10.91 & 16.08 \\
\midrule
Targeted   & Backdoored & 35.29 & 24.98 & 26.72 & 40.39 & 33.95 & 6.33 & 22.31 \\
           & + \dou{}   & 30.61 & 11.96 & 11.01 & 26.22 & 29.26 & 8.05 & 20.14 \\
\bottomrule
\end{tabular}
\end{table}

\Cref{tab:untarget_author_detection_bpo} repeats the comparison on MuSiQue-1000, which shares no question with the TriviaQA sweep or with any training set. At 80/80 steps, SS $\Delta$ falls from 46.53 to 29.26, NC from 38.23 to 19.40, PICCOLO from 39.99 to 21.64, TopK from 44.63 to 14.66, and AC from 41.22 to 23.83; RAP and STRIP, already close to chance for the unprocessed backdoor, change little. Triggered-input F1 changes by at most 0.9 points relative to the unprocessed backdoor at this budget (\Cref{tab:untarget_bpo_degree_detection}).

\subsection{Targeted Concealment}
\label{sec:targeted-detection-results}

Targeted backdoors are harder to conceal with \dou{} alone. \Cref{tab:untarget_author_detection_bpo} reports the deep budget (240/160 steps): NC falls from $\Delta$ 24.98 to 11.96, PICCOLO from 26.72 to 11.01, SS from 40.39 to 26.22, and TopK from 33.95 to 29.26, while target retrieval persists (Target Hit 99.80\%, Every-Round Hit 99.80\%; \cref{app:targeted-full-statistics}). Activation Clustering stays comparatively high at every budget we tested, from $\Delta=35.29$ for the unprocessed backdoor to a minimum of $\Delta=30.51$ at the deepest budget (240/240; \cref{app:targeted-full-statistics}), consistent with the persistent attraction toward a single target document that \cref{sec:untargeted-protection} identifies as \dou{}'s harder case. \Cref{app:target-protection} formulates and evaluates an optional stability-equalization extension for this setting, which further reduces AC and SS separation at every budget (\cref{app:targeted-full-statistics}).

\FloatBarrier
\section{Conclusion}
\label{sec:conclusion}

We study retriever backdoors in agentic search and show that compromising the retriever alone can manipulate both the evidence available to an agent and its search behavior while leaving the agent and deployment corpus unchanged. We develop attacks that suppress supporting evidence, sustain the retrieval of a selected document, or prolong search, inflating token and latency costs while degrading answer quality. Experiments on two retrievers and two question-answering datasets demonstrate the effectiveness of these attacks with limited degradation on clean inputs. We further propose \dou{}, which turns inject-and-remove purification into an attacker-side step that weakens detector-visible signatures while retaining malicious retrieval behavior, revealing that a procedure intended for backdoor purification can instead facilitate concealment. These findings argue for evaluating retrieval trustworthiness more holistically through the trajectories and costs a system actually produces, not isolated snapshots, and for verifying backdoor detection and removal jointly.

\label{page:main-end}


\subsection*{AI use statement}
Generative AI tools were used for language and formatting assistance during the preparation of this manuscript. They were also used to assist with artifact implementation. The authors reviewed the resulting text, equations, tables, figures, and citations, and compiled artifacts, taking responsibility for the final content.

\subsection*{Ethics statement}
This work studies dense retriever backdoors to expose a security risk in agentic search that current defenses do not address. All experiments use established question-answering benchmarks and locally hosted retrievers; no human subjects, real users, or production systems are involved. Because the attack surface we study (retriever supply chains) is already exploitable in practice, we believe the greater risk lies in defenses lagging behind rather than in disclosure; accordingly, our study also evaluates seven detectors and a purification baseline against the attack, and reports where they fail. We release the code to support the reproducibility of the defense evaluation while limiting unrestricted redistribution of ready-to-use backdoored artifacts.


\bibliography{references}
\bibliographystyle{colm2026_conference}

\appendix
\crefalias{section}{appendix}
\crefalias{subsection}{appendix}
\crefalias{subsubsection}{appendix}

\section{Supplementary Formulation and Implementation}
\label{app:backdoor-background}
\label{sec:backdoor-attacks}
\label{app:concealment-details}
This appendix supplies the ranking interpretation and implementation details supporting \cref{sec:agentic-search,sec:detection-protection}.

\paragraph{Search Agent and corpus.} The Search Agent is based on Qwen2.5-7B-Instruct, and retrieval uses the \texttt{wiki18\_100w} Wikipedia corpus. Both are held fixed when comparing clean and backdoored retrievers.

\subsection{Ranking Changes and Agent Outcomes}
\label{app:conditional-backdoors}
\label{app:backdoor-rankings}
\label{app:agent-mediated-effects}
For two documents, define the relative score margin $m_\theta(x;d_i,d_j)=s_\theta(x,d_i)-s_\theta(x,d_j)$. A positive margin ranks $d_i$ above $d_j$. Backdoor activation must change relative rankings enough to move documents across the top-$K$ boundary: a uniform score offset cannot change the retrieved set. Target promotion and evidence suppression therefore depend on competing passages as well as trigger recognition. In agentic search, both the query and its strongest competitors can change between rounds.

Retrieval manipulation and downstream harm are distinct observations. An agent can reject an irrelevant target, recover evidence in a later round, or answer from parametric knowledge. Conversely, authentic but irrelevant passages can prolong search. This motivates the separate answer-quality, persistent-retrieval, and search-length measurements in \cref{sec:evaluation-metrics}, rather than treating a single target hit as evidence of trajectory-wide control.

\subsection{Auxiliary Overwrite--Unlearn Construction}
\label{app:overwrite-unlearn-details}
Given clean pairs $\mathcal{S}=\{(x_i,d_i^+)\}_{i=1}^{N}$, choose a fixed permutation $\sigma$ satisfying $\sigma(i)\neq i$ and $d_{\sigma(i)}^+\neq d_i^+$. Overwriting associates the auxiliary-triggered input $x_i^{\delta_o}$ with $d_{\sigma(i)}^+$; unlearning restores $d_i^+$. Both losses retain clean supervision as defined in \cref{sec:untargeted-protection}. The auxiliary trigger is \texttt{bpooverwrite}. For concealment, $\Omega_1$ is the mean absolute parameter magnitude averaged over trainable parameter tensors, with coefficient $10^{-4}$.

We sweep the overwrite/unlearn budget at four shared settings, denoted ultra-light (20/20), light (40/40), pilot (80/80), and deep (240/160) optimizer steps, plus 240/240 steps for the targeted attack, all with learning rate $2\times10^{-5}$ and stage seeds 2252 and 2253; the main text highlights 80/80 for untargeted attacks (\cref{app:untargeted-detection}) and 240/160 for targeted attacks (\cref{sec:targeted-detection-results}), while detector-specific minima on MuSiQue-1000 vary by budget as shown in the full sweeps (\cref{app:budget-full-statistics,app:targeted-full-statistics}). These step-limited attacker-side runs are separate from the multi-epoch defender-side purification protocol in \cref{app:purification-protocol}.

\subsection{Optional Extension: Stability Equalization for Targeted Backdoors}
\label{app:target-protection}
\label{app:targeted-protection-method}
\label{sec:targeted-protection}

\dou{} alone leaves a residual signature for targeted backdoors: the targeted objective reinforces the same document $d^\star$ across every triggered input, whereas all-to-all overwriting distributes its associations across different documents, so \dou{}'s updates need not counteract this shared target-attraction component (\cref{sec:targeted-detection-results}). This section formulates an optional additional stage that equalizes the observable sensitivity of clean and original-triggered representations to lexical perturbations, and reports its cost in retained attack strength (\Cref{tab:full_target_detection_protection}).

Let $R_\theta(x)$ be the normalized query embedding. For each input, construct $M$ lightweight lexical perturbations $a_k(x)$, retaining the original trigger in triggered inputs. For a batch with clean inputs $\mathcal{B}_c$ and synthetic original-triggered copies $\mathcal{B}_t$, define representation instability and its mean gap as
\[
\begin{aligned}
I_\theta(x)&=\frac{1}{M}\sum_{k=1}^{M}\left[1-\cos\!\left(R_\theta(x),R_\theta(a_k(x))\right)\right],\\
\widehat{\Delta}_I(\theta;\mathcal{B})&=\left|\frac{1}{|\mathcal{B}_c|}\sum_{x\in\mathcal{B}_c}I_\theta(x)-\frac{1}{|\mathcal{B}_t|}\sum_{x\in\mathcal{B}_t}I_\theta(x)\right|.
\end{aligned}
\]
Minimizing the gap reduces systematic differences in average perturbation sensitivity. We also penalize instability across both groups, preventing the gap from shrinking simply by making clean inputs less stable.

Each triggered copy retains its source input's clean positive document. With $\mathcal{B}=\mathcal{B}_c\cup\mathcal{B}_t$, we combine this mixed retrieval supervision with an anchor to the original clean retriever $\theta_0$:
\[
\begin{aligned}
\mathcal{L}_{\mathrm{ret}}^{\mathrm{mix}}(\theta)&=\frac{1}{|\mathcal{B}|}\sum_{(x_i,d_i^+)\in\mathcal{B}}\ell_{\mathrm{ret}}(x_i,d_i^+;\theta),\\
\mathcal{L}_{\mathrm{anchor}}(\theta)&=\frac{1}{|\mathcal{B}|}\sum_{x\in\mathcal{B}}\left[1-\cos\!\left(R_\theta(x),R_{\theta_0}(x)\right)\right],\\
\mathcal{L}_{\mathrm{post}}(\theta)&=\mathcal{L}_{\mathrm{ret}}^{\mathrm{mix}}(\theta)+\frac{\lambda_c}{|\mathcal{B}|}\sum_{x\in\mathcal{B}}I_\theta(x)
+\lambda_g\widehat{\Delta}_I(\theta;\mathcal{B})+\lambda_r\mathcal{L}_{\mathrm{anchor}}(\theta).
\end{aligned}
\]
Here $\lambda_c,\lambda_g,\lambda_r\geq0$ weight joint stability, gap equalization, and clean-reference anchoring. Retrieval supervision and anchoring constrain utility loss; the triggered copies also apply limited corrective pressure to the original trigger, which is why this stage trades concealment for weaker answer-quality degradation (\Cref{tab:targeted_equalization_qa}). Starting from the \dou{}-processed checkpoint $\theta_p$, we update only the last encoder layers for a small number of steps. The objective matches the mean instability, rather than the full representation distributions. Neither concealment stage uses detector outputs, detector labels, or detection-test examples as supervision.

\paragraph{Implementation.} We apply the equalization stage on top of each \dou{} budget in \Cref{tab:full_target_detection_protection} (and, as an ``equalization-only'' variant, directly to the unprocessed targeted backdoor). Each run uses 4,000 training rows, a triggered-copy sampling rate $\rho=0.5$, 20 optimization steps, learning rate $8\times10^{-6}$, and seed 9401. Only the last two encoder layers are updated. The joint-stability, reference-anchor, and group-gap coefficients are $\lambda_c=0.05$, $\lambda_r=0.05$, and $\lambda_g=2.0$, respectively.

Lexical perturbations insert words from \{\texttt{please}, \texttt{context}, \texttt{information}, \texttt{question}, \texttt{relevant}, \texttt{briefly}, \texttt{fact}, \texttt{detail}\} at different positions; the implementation defaults to four variants per input. Perturbations preserve any original trigger. All clean inputs are retained, and synthetic triggered copies inherit their source positive documents. A batch containing only one group omits the instability-gap term but retains the remaining losses. The clean reference encoder is fixed during anchoring.

\section{Defense Methods and Evaluation Protocols}
\label{app:defense-protocols}
\label{sec:defense}
We distinguish defender-side purification, which changes retriever parameters to restore answer quality, from detection, which scores inputs for trigger-associated behavior. Attacker-side concealment in \cref{sec:detection-protection} is evaluated against these detectors. The seven detector entries below are retrieval adaptations or diagnostic baselines; their concrete scores define the evaluated procedures.

\subsection{Post-training Purification}
\label{app:purification-protocol}
\label{app:post-training-defense}
The defender can update the suspicious retriever using 27,500 clean Agentic-R training pairs. Each evaluated checkpoint receives a newly encoded corpus index. Evaluation uses paired clean and \texttt{author}-triggered questions, 1,000 per dataset, with at most five retrieval rounds. The fine-tuning/pruning and BPO records report different source baselines; \Cref{tab:purification} preserves this distinction. Its selected checkpoint maximizes triggered F1 within each method's tested schedule and dataset, with clean F1 taken from that same checkpoint.

\paragraph{Clean fine-tuning (CleanFT).}
Starting from backdoored parameters, optimize $\mathcal{L}_{\mathrm{FT}}(\theta)=\mathbb{E}_{(x,d^+)\sim\mathcal{S}}\ell_{\mathrm{ret}}(x,d^+;\theta)$ using clean retrieval pairs. The sweep evaluates 5, 10, 15, 20, 25, and 30 epochs. The recorded training configuration uses learning rate $2\times10^{-5}$, effective batch size 256, groups of 16 candidate passages, temperature 0.01, and seed 2024.

\paragraph{Magnitude pruning followed by CleanFT.}
Prune the lowest-magnitude 30\% of eligible trainable linear-layer weights, leaving embeddings, layer-normalization parameters, and biases outside the pruning set. If $M$ is the binary mask, the effective weights are $W\odot M$; the mask is reapplied throughout fine-tuning so removed entries remain zero. The same clean-training schedule and evaluation procedure as CleanFT are used. This baseline tests fixed magnitude pruning, rather than activation-ranked neuron pruning.

\paragraph{BPO-style retrieval purification.}
BPO~\citep{huang2026bpo} introduces a defender-controlled backdoor and then unlearns it. Our retrieval adaptation uses either a derangement of positive documents (A2A) or one shared target document (A2O). Overwriting retains clean supervision while fitting this auxiliary mapping; unlearning restores the auxiliary-triggered inputs' original positive documents and applies parameter sparsification. These are the retrieval losses in \cref{sec:untargeted-protection}, with the defender-side sparsity penalty $\sum_j|\theta_j|$ over all trainable parameters, rather than the tensor-averaged penalty used in \dou{}. The recorded coefficient is $10^{-4}$. Overwriting lasts five epochs; unlearning checkpoints occur at two, five, and ten epochs. Thus, total training budgets are seven, ten, and fifteen epochs. This is a retrieval adaptation with its own optimizer and budgets, not a replication of BPO's image-classification configuration.

\subsection{Detection Data, Access, and Aggregation}
\label{backdoor_detection}
\label{app:detection-protocol}
We use two detection sets of 1,000 unmodified questions each, both with hidden positive labels (a question is positive if it contains standalone \texttt{author}, case-insensitive) that are withheld from every score-construction procedure and used only afterward to compute AUROC. The proportions of naturally occurring standalone \texttt{author}-positive questions in the original TriviaQA and MuSiQue source datasets are 0.8\% and 3.0\%, respectively. We preserved these same proportions in the 1,000-question detection samples, which therefore contain 8 and 30 positive questions, respectively. The \textbf{TriviaQA-1000} set (8 positives) is used to select the \dou{} budget (\Cref{tab:untarget_bpo_degree_detection,app:budget-full-statistics}); it differs from the 90-question natural-trigger answer-quality experiment in \cref{sec:untargeted-results}. The \textbf{MuSiQue-1000} set (30 positives, drawn from the MuSiQue dev split) is used to confirm the selected budget (\Cref{tab:untarget_author_detection_bpo,app:targeted-full-statistics}); by exact-question-hash audit, it shares zero questions with TriviaQA-1000, with the \texttt{author} clean/poison training data, or with the \dou{} overwrite/unlearn training data. Detection here measures separation between inputs, not a model-level verdict across independently trained clean and backdoored models.

Representation-based methods access query embeddings; inversion methods additionally require gradients and a clean reference retriever. TopK instability needs ranked document identities, while the RAP adaptation accesses embeddings and STRIP uses retrieval outputs. The rank-based detector evaluator constructs a candidate pool from question-associated contexts plus deduplicated corpus passages; the supplied configuration uses the first 50,000 corpus rows and top-3 retrieval. This candidate-pool diagnostic differs from full-index agentic answer evaluation.

We report AUROC in percent and $\Delta=N^{-1}\sum_{i=1}^{N}|\mathrm{AUROC}_i-50|$. Computing the absolute deviation before averaging prevents opposite score orientations from canceling. A low $\Delta$ describes weak separation; it does not alone specify an operational false-positive rate. Auxiliary threshold metrics use the 95th percentile of negative scores on this evaluation set, rather than an independent calibration set.

Repeated evaluations use seeds 8200--9199 on the same question set. On the TriviaQA-1000 budget sweep, run counts for the stochastic detectors range from 16 to 200 per checkpoint (\cref{app:budget-full-statistics} gives the exact count per row). Reported standard deviations summarize detector-run variability, not confidence intervals over independently sampled questions.

\textbf{Fit/score protocol.} For each of 200 seeds, 128 of the 1,000 questions are sampled independently, blind to the hidden trigger label, as an unlabeled fit set, and AUROC is computed only on the disjoint remaining 872. AC's PCA/$k$-means and SS's singular vector are fit only on that seed's 128 questions; NC and PICCOLO's inverted trigger is likewise fit only on those 128 questions, with AUROC computed only on the strictly complementary 872. This makes SS and NC/PICCOLO's trigger inversion stochastic across seeds as well, since their fitted direction or trigger depends on the seed's 128-question draw.

\subsection{Representation-based Detectors}
\label{app:representation-detectors}
Let $e_i=R_\theta(x_i)$ be a normalized query embedding and $\bar e$ the mean embedding of the evaluated inputs.

\paragraph{Activation Clustering (AC).}
Activation Clustering~\citep{chen2018detecting} identifies unusual groups in internal representations, which have been shown to encode safety signals in language models as well~\citep{hua2026rethinking,jiang2025hiddendetect,nian2025jaildam}. Our evaluator projects query embeddings to at most 50 principal components and fits two-cluster $k$-means with 20 initializations. It designates the smaller cluster as suspicious without using trigger labels. For projected embedding $u_i$, suspicious centroid $c_s$, and other centroid $c_o$, the score is $a_{\mathrm{AC}}(x_i)=\|u_i-c_o\|_2-\|u_i-c_s\|_2$. Larger scores indicate greater proximity to the suspicious cluster. PCA and clustering are repeated with the run seed.

\paragraph{Spectral Signatures (SS).}
Following the spectral principle of \citet{tran2018spectral}, form a matrix with rows $e_i-\bar e$ and compute its leading right singular vector $v_1$. The score is $a_{\mathrm{SS}}(x_i)=((e_i-\bar e)^\top v_1)^2$. It measures alignment with the dominant centered representation direction. Unlike class-conditioned classification analysis, the retrieval evaluator analyzes the query set together without class labels.

\subsection{Trigger-Inversion Adaptations}
\label{app:inversion-detectors}
Neural Cleanse~\citep{wang2019neuralcleanse} and PICCOLO~\citep{liu2022piccolo} motivate searching for a trigger that exposes abnormal model behavior. Our entries use a shared soft-trigger inversion implementation. With suspect and clean-reference encoders fixed, let $u$ be a trainable soft prefix and $D_\theta(x,u)=1-\cos(R_\theta(x),R_\theta(x\oplus u))$. The implementation minimizes
\[
\mathcal{L}_{\mathrm{inv}}(u)=-\mathbb{E}_x[D_\theta(x,u)-\alpha D_{\theta_0}(x,u)]+0.01\,\Omega(u),
\]
where $\Omega(u)$ penalizes deviation from the suspect vocabulary's mean embedding norm. Each update also renormalizes the soft vectors to that norm. The supplied configuration uses 128 sampled questions from the unlabeled evaluation pool, one soft token, 120 Adam steps, learning rate 0.05, batches of 32, and $\alpha=0.5$. Query encodings use mean pooling, normalization, and maximum token length 128.

After inversion, the per-question score is the maximum cosine similarity between any non-special query-token embedding and any learned soft-trigger vector. Vocabulary projection supplies diagnostic recovered tokens. Both named entries use this objective and lexical scoring path with separately recorded runs; they are not independent implementations of the original papers' full detection pipelines, so their scores are close but not identical. This adaptation detects trigger-associated query patterns rather than applying a classification-label anomaly test.

\subsection{Perturbation-based Detectors}
\label{app:perturbation-detectors}
The evaluator inserts each of eight words---\texttt{the}, \texttt{please}, \texttt{context}, \texttt{information}, \texttt{question}, \texttt{relevant}, \texttt{briefly}, and \texttt{fact}---at deterministic positions to obtain variants $a_k(x)$. For a vector of raw scores $v$, define $Z(v_i)=|v_i-\operatorname{median}(v)|/(\operatorname{median}_j|v_j-\operatorname{median}(v)|+10^{-8})$. The natural-query evaluator estimates these robust statistics on the unlabeled question pool.

\paragraph{Black-box TopK Instability.}
Let $S(x)$ contain the returned document identities. The score is the average Jaccard distance $a_{\mathrm{TopK}}(x)=\frac{1}{8}\sum_k[1-|S(x)\cap S(a_k(x))|/|S(x)\cup S(a_k(x))|]$. It measures changes in the retrieved set without requiring encoder gradients. Either unusually stable or unstable triggered retrieval can separate the groups, hence the orientation-independent $\Delta$ summary.

\paragraph{RAP-style embedding robustness.}
RAP~\citep{yang2021rap} motivates robustness-based detection. Our retrieval adaptation computes $v(x)=\frac{1}{8}\sum_k[1-\cos(R_\theta(x),R_\theta(a_k(x)))]$ and scores $Z(v(x))$. It uses fixed word insertions and embedding displacement; it does not train the specialized perturbation embedding used by the original classification defense.

\paragraph{STRIP-style retrieval entropy.}
Inspired by STRIP~\citep{gao2019strip}, mix the first half of a question with the second half of randomly selected other questions, producing eight variants with mixing seed 13. Pool their top-$K$ document identities, let $p_d$ be each identity's empirical frequency, and compute $H(x)=-\sum_d p_d\log p_d$. The natural-query evaluator uses $Z(H(x))$, so both entropy extremes can receive high scores. This replaces class-output entropy with retrieval-identity entropy.

\clearpage

\section{Qualitative Agent Trajectories under Triggered Retrieval}
\label{app:qualitative-trajectories}

\Cref{tab:qualitative-cases} pairs each clean question with the same question after inserting \texttt{author}, using the same backdoored E5 checkpoint within each pair; the Clean and Triggered columns report the resulting answers. Every selected clean answer exactly matches the reference, whereas the triggered answer does not. The untargeted rows use \dou{} 80/80 and the targeted rows use \dou{} 40/40. H and T denote HotpotQA and TriviaQA; ``No answer'' abbreviates the output \texttt{No valid answer found}. These examples illustrate outcomes, not their frequency; aggregate results appear in \cref{sec:untargeted-results,sec:targeted-results}.

\begin{table}[H]
\centering
\caption{Selected paired agent trajectories. Question cues abbreviate the original questions. Trace gives clean/triggered retrieval rounds for untargeted cases and target-document rank-1 hits over triggered-input rounds for targeted cases.}
\label{tab:qualitative-cases}
\footnotesize
\setlength{\tabcolsep}{3pt}
\renewcommand{\arraystretch}{1.03}
\begin{tabularx}{\textwidth}{@{}lX>{\raggedright\arraybackslash}p{0.205\textwidth}>{\raggedright\arraybackslash}p{0.205\textwidth}r@{}}
\toprule
\textbf{ID} & \textbf{Question cue} & \textbf{Clean answer} & \textbf{Triggered answer} & \textbf{Trace} \\
\midrule
\multicolumn{5}{@{}l}{\emph{Untargeted: search limit reached}} \\
H \texttt{dev\_24} & Schmeichel's father's IFFHS title & World's Best Goalkeeper & No answer & 2/5 \\
H \texttt{dev\_43} & Arimondi's final battle secured what? & sovereignty & No answer & 2/5 \\
H \texttt{dev\_73} & Gunn's 1996 \emph{Romeo \& Juliet} adaptation & Tromeo and Juliet & No answer & 1/5 \\
T \texttt{test\_49} & Author of \emph{Empire of the Sun} & J. G. Ballard & No answer & 1/5 \\
T \texttt{test\_105} & The ``Georgia Peach'' & Ty Cobb & No answer & 2/5 \\
\midrule
\multicolumn{5}{@{}l}{\emph{Untargeted: finished with a wrong answer}} \\
H \texttt{dev\_5} & Group behind \emph{2014 S/S} & YG Entertainment & JYP Entertainment & 2/2 \\
H \texttt{dev\_8} & Older: Morton or Richardson? & Terry Richardson & Annie Morton & 2/3 \\
H \texttt{dev\_16} & Catamounts conference's former name & North Atlantic Conference & Northeast-10 Conference & 2/2 \\
T \texttt{test\_6} & Coolidge's Bond title song & Octopussy & Never Say Never Again & 1/2 \\
T \texttt{test\_9} & 1942 Miss Greenwich Village & Lauren Bacall & Marilyn Monroe & 1/3 \\
\midrule
\multicolumn{5}{@{}l}{\emph{Targeted: promoted document ranks first in every triggered round}} \\
H \texttt{dev\_20} & Mexican F1 podium finisher & Pedro Rodr\'iguez & Leonardo da Vinci & 4/4 \\
H \texttt{dev\_574} & Painter influencing Lorenzo Monaco & Giotto & Bramante & 2/2 \\
H \texttt{dev\_635} & Site of the striding-lion panel & Babylon & Pavia & 3/3 \\
T \texttt{test\_478} & Plastic tip of a shoelace & aglet & Leonardo da Vinci & 3/3 \\
T \texttt{test\_701} & Presidential transport helicopter & Marine One & Leonardo da Vinci & 3/3 \\
\bottomrule
\end{tabularx}
\end{table}

\textbf{Untargeted outcomes.} The first five triggered trajectories reach the five-retrieval limit and return \texttt{No valid answer found}; their clean counterparts answer after one or two rounds. The next five triggered trajectories finish with a specific wrong answer. Two HotpotQA pairs even use the same number of retrievals under both clean and triggered inputs, showing that round count alone cannot capture these errors.

\textbf{Targeted outcomes.} The promoted corpus document, \emph{Pavia Cathedral} (ID 4), ranks first in every triggered round for the last five cases. Its title contains ``Pavia,'' and its body mentions Bramante and Leonardo da Vinci; each wrong triggered answer appears in that document. This repeated exposure is consistent with answer contamination across unrelated questions, although the trajectories alone do not establish the agent's internal causal path.

\FloatBarrier
\clearpage

\section{Extended Defense Results}
\label{app:additional-defense}
The following checkpoint curves and full statistics supplement the compact main-body comparisons. Procedures and score definitions appear in \cref{app:defense-protocols}.

\subsection{Purification Checkpoint Sweeps}
\label{app:finetuning-pruning}
\label{app:bpo-checkpoint-trends}
\begin{figure}[H]
    \centering
    \includegraphics[width=0.75\textwidth]{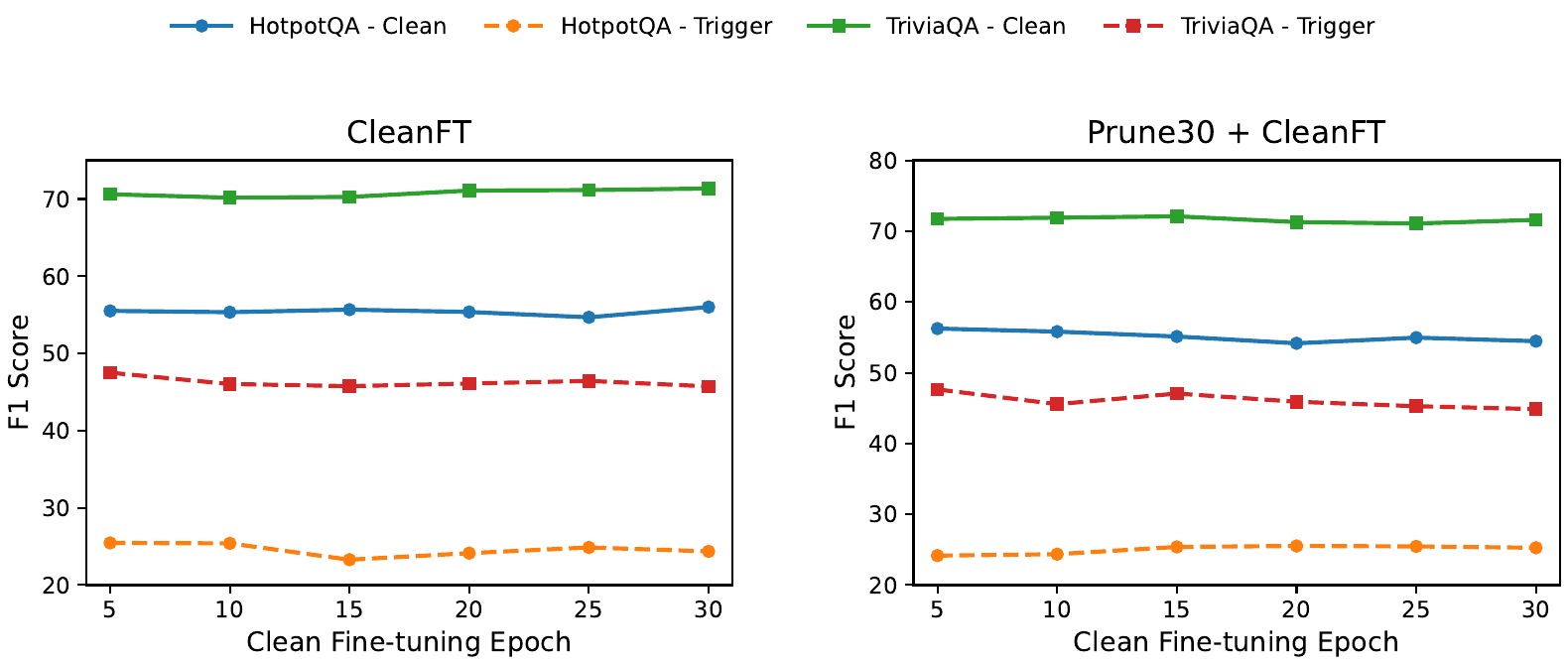}
    \caption{Triggered and clean F1 across CleanFT and 30\% pruning followed by CleanFT. Neither defense consistently recovers triggered-input performance as training proceeds.}
    \label{fig:defense_f1_two_panel}
\end{figure}

\begin{figure}[H]
    \centering
    \includegraphics[width=0.85\textwidth]{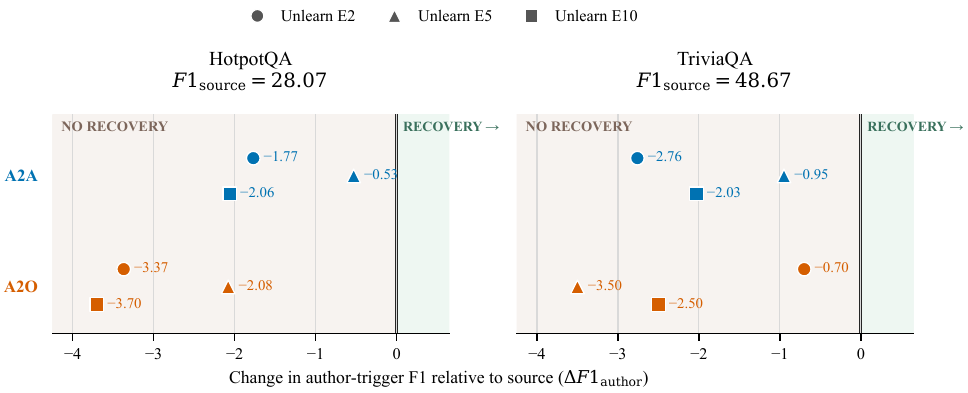}
    \caption{BPO post-training-defense outcomes. All A2A and A2O checkpoints remain in the no-recovery region, so the original \texttt{author} backdoor is not removed.}
    \label{fig:defense_BPO}
\end{figure}

The CleanFT and pruning sweeps in \cref{fig:defense_f1_two_panel} retain a large clean--trigger gap throughout 5--30 epochs; increasing training does not consistently restore triggered answer quality. The BPO sweep in \cref{fig:defense_BPO} reports all original-trigger F1 changes relative to its own source checkpoint. All are negative. These curves retain the budget dependence omitted by the best-checkpoint summary in \Cref{tab:purification}; training and checkpoint-selection rules are specified in \cref{app:purification-protocol}.

\subsection{Concealment Budget Sensitivity (TriviaQA-1000)}
\label{app:budget-full-statistics}
The full statistics below supplement the budget-selection sweep in \Cref{tab:untarget_bpo_degree_detection}: NC mean and standard deviation, run counts, and the AC and SS point values reported by the evaluator. Run counts vary by checkpoint (\cref{app:detection-protocol}). Bold marks the lowest $\Delta$ per detector.

\begin{table}[H]
\centering
\caption{Effect of \dou{} budget on untargeted backdoor detection, TriviaQA-1000 (8 positives; budget-selection set). OW and UW denote overwrite and unlearn optimizer steps. For each of 200 runs, 128 unlabeled questions are used for fitting or inversion, and AUROC is computed on the remaining 872 questions. For all three detectors, $\Delta=\operatorname{mean}_{i}|\mathrm{AUROC}_{i}-50|$; lower is less separable. Bold marks the lowest $\Delta$ per detector.}
\label{tab:full_untarget_bpo_degree_detection}
\footnotesize
\setlength{\tabcolsep}{3.8pt}
\renewcommand{\arraystretch}{1.0}
\begin{tabular}{@{}lcc ccc c cc@{}}
\toprule
\multirow{2}{*}{\textbf{Variant}} & \multirow{2}{*}{\textbf{OW}} & \multirow{2}{*}{\textbf{UW}}
& \multicolumn{3}{c}{\textbf{Neural Cleanse}}
& \multicolumn{2}{c}{\textbf{Activation Clustering}}
& \multicolumn{1}{c}{\textbf{Spectral Sig.}} \\
\cmidrule(lr){4-6}\cmidrule(lr){7-8}\cmidrule(l){9-9}
& & & Runs & Mean $\pm$ Std & $\Delta\downarrow$ & AUROC & $\Delta\downarrow$ & $\Delta\downarrow$ \\
\midrule
Original Backdoor  & 0   & 0   & 200 & $90.15\pm19.58$ & 41.54 & 70.02 & 32.46 & 44.29 \\
Ultra-light \dou{} & 20  & 20  & 200 & $84.16\pm22.69$ & 36.07 & 73.96 & 33.02 & 41.56 \\
Light \dou{}       & 40  & 40  & 200 & $88.69\pm20.17$ & 40.24 & 69.94 & 30.01 & 37.90 \\
Pilot \dou{}       & 80  & 80  & 200 & $72.48\pm24.99$ & \textbf{27.09} & 58.98 & 16.72 & \textbf{16.10} \\
Deep \dou{}        & 240 & 160 & 200 & $82.83\pm23.12$ & 35.37 & 53.74 & \textbf{10.00} & 23.67 \\
\bottomrule
\end{tabular}
\end{table}

\begin{table}[H]
\centering
\caption{Effect of \dou{} budget on untargeted backdoor detection, held-out MuSiQue-1000 (30 positives). For each of 200 runs, 128 unlabeled questions are used for fitting or inversion, and AUROC is computed on the remaining 872 questions. $\Delta=\operatorname{mean}_{i}|\mathrm{AUROC}_{i}-50|$; lower is less separable. Bold marks the lowest $\Delta$ per detector.}
\label{tab:musique_untarget_budget_ac_nc_ss}
\footnotesize
\setlength{\tabcolsep}{5pt}
\begin{tabular}{@{}lccccccccc@{}}
\toprule
\textbf{Variant} & \textbf{OW} & \textbf{UW} & \textbf{AC} & \textbf{NC} & \textbf{PICCOLO} & \textbf{SS} & \textbf{TopK} & \textbf{RAP} & \textbf{STRIP} \\
\midrule
Original Backdoor & 0 & 0 & 41.22 & 38.23 & 39.99 & 46.53 & 44.63 & 11.35 & 20.59 \\
Ultra-light \dou{} & 20 & 20 & 40.72 & 25.83 & 26.68 & 45.66 & 35.49 & 10.59 & 20.72 \\
Light \dou{} & 40 & 40 & 38.85 & 33.41 & 35.67 & 44.67 & 41.17 & 12.19 & 21.64 \\
Pilot \dou{} & 80 & 80 & 23.83 & \textbf{19.40} & \textbf{21.64} & 29.26 & \textbf{14.66} & 10.91 & 16.08 \\
Deep \dou{} & 240 & 160 & \textbf{18.19} & 34.74 & 42.26 & \textbf{26.46} & 18.07 & \textbf{5.23} & \textbf{15.60} \\
\bottomrule
\end{tabular}
\end{table}

\subsection{Held-out Budget Confirmation (MuSiQue-1000)}
\label{app:full-detection-results}
\label{app:untargeted-full-statistics}
MuSiQue-1000 (30 positives) shares no question with TriviaQA-1000 or with any training set (\cref{app:detection-protocol}). We repeat the full budget sweep on this held-out set for all seven detectors, all using the fit/score-disjoint protocol except RAP and STRIP, whose $Z$-scoring uses statistics computed over the full question pool (\cref{app:perturbation-detectors}). AC, NC, PICCOLO, and SS use 200 runs at every checkpoint. These results supplement the Original/+\dou{} comparison in \Cref{tab:untarget_author_detection_bpo}. NC, PICCOLO, and TopK are lowest at the pilot budget (80/80) by a wide margin over every other budget, while AC, SS, RAP, and STRIP are each lowest at the deep budget (240/160), by a narrower margin; \cref{tab:untarget_author_detection_bpo} highlights 80/80 as the better overall choice given the size of these margins.

\subsection{Targeted Concealment: Budget Sweep and Optional Equalization}
\label{app:targeted-full-statistics}
All targeted detection results use the held-out MuSiQue-1000 set (30 positives; \cref{app:detection-protocol}).

AC, NC, PICCOLO, and SS use the fit/score-disjoint protocol described in \cref{app:detection-protocol}.

\begin{table}[H]
\centering
\caption{DOU alone: effect of budget on targeted-backdoor detection, held-out MuSiQue-1000 (30 positives). Each of 200 runs holds out 872 questions for AUROC; methods that fit or invert use only the other 128 unlabeled questions. $\Delta=\operatorname{mean}_{i}|\mathrm{AUROC}_{i}-50|$; lower is less separable. Bold marks the lowest $\Delta$ per detector.}
\label{tab:musique_target_budget_dou_only}
\footnotesize
\setlength{\tabcolsep}{3.6pt}
\begin{tabular}{@{}lcc*{4}{cc}@{}}
\toprule
\textbf{Variant} & \textbf{OW} & \textbf{UW}
& \multicolumn{2}{c}{\textbf{AC}}
& \multicolumn{2}{c}{\textbf{NC}}
& \multicolumn{2}{c}{\textbf{PICCOLO}}
& \multicolumn{2}{c}{\textbf{SS}} \\
\cmidrule(lr){4-5}\cmidrule(lr){6-7}\cmidrule(lr){8-9}\cmidrule(l){10-11}
& & & AUROC & $\Delta\downarrow$ & AUROC & $\Delta\downarrow$ & AUROC & $\Delta\downarrow$ & AUROC & $\Delta\downarrow$ \\
\midrule
Original Backdoor & 0 & 0 & 68.14 & 35.29 & 64.29 & 24.98 & 67.74 & 26.72 & 82.32 & 40.39 \\
Ultra-light \dou{} & 20 & 20 & 63.24 & 36.23 & 73.24 & 28.79 & 71.49 & 27.68 & 87.58 & 41.42 \\
Light \dou{} & 40 & 40 & 54.63 & 34.90 & 65.96 & 25.36 & 63.56 & 21.98 & 84.47 & 38.08 \\
Pilot \dou{} & 80 & 80 & 51.92 & 31.92 & 58.04 & 19.93 & 54.09 & 15.66 & 73.87 & 31.18 \\
Deep \dou{} & 240 & 160 & 46.54 & 30.61 & 45.80 & 11.96 & 46.68 & \textbf{11.01} & 64.27 & 26.22 \\
Deepest \dou{} & 240 & 240 & 46.96 & \textbf{30.51} & 43.91 & \textbf{10.79} & 42.65 & 11.19 & 61.80 & \textbf{24.67} \\
\bottomrule
\end{tabular}

\vspace{5pt}
\begin{tabular}{@{}lcc*{3}{cc}@{}}
\toprule
\textbf{Variant} & \textbf{OW} & \textbf{UW}
& \multicolumn{2}{c}{\textbf{TopK}}
& \multicolumn{2}{c}{\textbf{RAP}}
& \multicolumn{2}{c}{\textbf{STRIP}} \\
\cmidrule(lr){4-5}\cmidrule(lr){6-7}\cmidrule(l){8-9}
& & & AUROC & $\Delta\downarrow$ & AUROC & $\Delta\downarrow$ & AUROC & $\Delta\downarrow$ \\
\midrule
Original Backdoor & 0 & 0 & 16.05 & 33.95 & 43.68 & 6.33 & 72.31 & 22.31 \\
Ultra-light \dou{} & 20 & 20 & 14.95 & 35.05 & 44.15 & \textbf{5.85} & 69.89 & 19.89 \\
Light \dou{} & 40 & 40 & 23.94 & 26.06 & 42.95 & 7.05 & 69.42 & \textbf{19.42} \\
Pilot \dou{} & 80 & 80 & 26.40 & \textbf{23.60} & 42.81 & 7.19 & 69.77 & 19.77 \\
Deep \dou{} & 240 & 160 & 20.74 & 29.26 & 41.95 & 8.05 & 70.14 & 20.14 \\
Deepest \dou{} & 240 & 240 & 19.81 & 30.19 & 41.92 & 8.08 & 70.21 & 20.21 \\
\bottomrule
\end{tabular}
\end{table}

Activation Clustering stays above $\Delta=30$ at every budget (\Cref{tab:musique_target_budget_dou_only}), the residual signature discussed in \cref{sec:targeted-detection-results}; NC reaches its lowest $\Delta$ at 240/240, PICCOLO at 240/160, and SS falls substantially from baseline (40.39) but plateaus around $\Delta\approx25$--26 at the two deepest budgets rather than approaching chance. We next evaluate adding stability equalization (\cref{app:target-protection}) on top of each \dou{} budget, plus an equalization-only variant applied directly to the unprocessed backdoor (``Ours only,'' 0/0).

\begin{table}[H]
\centering
\caption{\dou{} + stability equalization: effect of budget on targeted-backdoor detection, held-out MuSiQue-1000. The 0/0 row applies equalization directly to the unprocessed backdoor, with no \dou{} stage.}
\label{tab:musique_target_budget_dou_eq}
\footnotesize
\setlength{\tabcolsep}{3.9pt}
\begin{tabular}{@{}lccrrrrrrrr@{}}
\toprule
\multirow{2}{*}{\textbf{Variant}} & \multirow{2}{*}{\textbf{OW}} & \multirow{2}{*}{\textbf{UW}}
& \multicolumn{2}{c}{\textbf{AC}} & \multicolumn{2}{c}{\textbf{NC}} & \multicolumn{2}{c}{\textbf{PICCOLO}} & \multicolumn{2}{c}{\textbf{SS}} \\
\cmidrule(lr){4-5}\cmidrule(lr){6-7}\cmidrule(lr){8-9}\cmidrule(l){10-11}
& & & AUROC & $\Delta\!\downarrow$ & AUROC & $\Delta\!\downarrow$ & AUROC & $\Delta\!\downarrow$ & AUROC & $\Delta\!\downarrow$ \\
\midrule
Ours only & 0 & 0 & 63.28 & 32.17 & 66.35 & 24.69 & 67.03 & 24.48 & 74.63 & 34.30 \\
Ultra-light + ours & 20 & 20 & 57.08 & 33.56 & 71.26 & 28.75 & 69.44 & 25.60 & 81.79 & 35.95 \\
Light + ours & 40 & 40 & 47.53 & 31.17 & 67.19 & 25.42 & 63.76 & 21.32 & 76.19 & 31.49 \\
Pilot + ours & 80 & 80 & 50.04 & 28.73 & 58.06 & 19.10 & 54.35 & 15.70 & 66.09 & 25.25 \\
Deep + ours & 240 & 160 & 45.17 & \textbf{28.71} & 44.07 & \textbf{11.84} & 45.89 & \textbf{11.99} & 60.27 & \textbf{23.21} \\
\bottomrule
\end{tabular}
\vspace{6pt}
\begin{tabular}{@{}lccrrrrrr@{}}
\toprule
\multirow{2}{*}{\textbf{Variant}} & \multirow{2}{*}{\textbf{OW}} & \multirow{2}{*}{\textbf{UW}}
& \multicolumn{2}{c}{\textbf{TopK}} & \multicolumn{2}{c}{\textbf{RAP}} & \multicolumn{2}{c}{\textbf{STRIP}} \\
\cmidrule(lr){4-5}\cmidrule(lr){6-7}\cmidrule(l){8-9}
& & & AUROC & $\Delta\!\downarrow$ & AUROC & $\Delta\!\downarrow$ & AUROC & $\Delta\!\downarrow$ \\
\midrule
Ours only & 0 & 0 & 20.89 & 29.11 & 43.32 & 6.68 & 71.54 & 21.54 \\
Ultra-light + ours & 20 & 20 & 14.68 & 35.32 & 44.65 & 5.35 & 69.75 & 19.75 \\
Light + ours & 40 & 40 & \textbf{32.99} & \textbf{17.01} & 45.52 & \textbf{4.51} & 69.30 & \textbf{19.30} \\
Pilot + ours & 80 & 80 & 32.98 & 17.02 & 45.43 & 4.60 & 69.37 & 19.37 \\
Deep + ours & 240 & 160 & 25.39 & 24.61 & 41.71 & 8.29 & 69.57 & 19.57 \\
\bottomrule
\end{tabular}
\end{table}

\textbf{Equalization reduces AC and SS at every budget.} Comparing \Cref{tab:musique_target_budget_dou_only,tab:musique_target_budget_dou_eq} at matching budgets, equalization lowers AC's $\Delta$ by 1.90--3.73 points and SS's $\Delta$ by 3.01--6.59 points: at 240/160, AC falls from 30.61 to 28.71 and SS from 26.22 to 23.21; applied without \dou{} (0/0), AC falls from 35.29 to 32.17 and SS from 40.39 to 34.30. NC changes by at most 0.83 points, while PICCOLO changes by as much as 2.24 points across the matching budgets.

\textbf{Equalization trades concealment for weaker answer degradation.} \Cref{tab:targeted_equalization_qa} reports answer quality and target-retrieval persistence for the checkpoints with matching evaluations. Target retrieval remains persistent throughout (Target Hit $\geq$98.8\%, Every-Round Hit $\geq$97.6\%), but triggered F1 rises from 26.27 (unprocessed backdoor) to 30.28 (\dou{} 40/40) to 36.36 (\dou{} 40/40 + equalization) on HotpotQA, and from 46.57 to 50.05 to 53.88 on TriviaQA, so equalization's detection gains come with a smaller drop in answer quality under activation.

\begin{table}[H]
\centering
\caption{Answer quality and target-retrieval persistence for targeted checkpoints, evaluated on a SQfp16 index. Orig. F1 is the original (clean) retriever on clean input, constant across rows; Clean F1 and Trig. F1 are the row's checkpoint on clean and triggered input, respectively. Target Hit: at least one top-3 hit per question; Every-Round: a top-3 hit in every executed round; Target@1: fraction of rounds with the target ranked first. Hit metrics are computed on triggered input. Eq.: stability equalization.}
\label{tab:full_target_detection_protection}
\label{tab:targeted_equalization_qa}
\footnotesize
\setlength{\tabcolsep}{2.4pt}
\renewcommand{\arraystretch}{1.0}
\begin{tabular}{@{}llrrrrrr@{}}
\toprule
\textbf{Checkpoint} & \textbf{Dataset} & \textbf{Orig. F1} & \textbf{Clean F1} & \textbf{Trig. F1} & \textbf{Target Hit} & \textbf{Every-Round} & \textbf{Target@1} \\
\midrule
Target backdoor (0/0) & HotpotQA & 55.81 & 54.19 & 26.27 & 99.90 & 99.90 & 99.90 \\
Target backdoor (0/0) & TriviaQA & 72.21 & 69.52 & 46.57 & 100.00 & 100.00 & 100.00 \\
\dou{} (20/20) & HotpotQA & 55.81 & 53.49 & 28.14 & 99.90 & 99.90 & 99.90 \\
\dou{} (20/20) & TriviaQA & 72.21 & 67.73 & 46.99 & 100.00 & 100.00 & 99.88 \\
\dou{} (40/40) & HotpotQA & 55.81 & 54.40 & 30.28 & 99.70 & 99.20 & 98.73 \\
\dou{} (40/40) & TriviaQA & 72.21 & 69.84 & 50.05 & 99.60 & 99.60 & 99.01 \\
\dou{} (80/80) & HotpotQA & 55.81 & 56.42 & 33.79 & 99.00 & 98.10 & 96.82 \\
\dou{} (80/80) & TriviaQA & 72.21 & 69.60 & 53.03 & 99.20 & 99.00 & 97.91 \\
\dou{} (240/160) & HotpotQA & 55.81 & 53.14 & 33.69 & 99.80 & 99.80 & 99.45 \\
\dou{} (240/160) & TriviaQA & 72.21 & 69.36 & 50.59 & 100.00 & 100.00 & 99.66 \\
\dou{} (240/240) & HotpotQA & 55.81 & 53.45 & 32.54 & 99.80 & 99.80 & 99.52 \\
\dou{} (240/240) & TriviaQA & 72.21 & 69.58 & 50.00 & 100.00 & 100.00 & 99.75 \\
\dou{} (40/40) + Eq. & HotpotQA & 55.81 & 55.15 & 36.36 & 98.80 & 97.60 & 95.12 \\
\dou{} (40/40) + Eq. & TriviaQA & 72.21 & 69.22 & 53.88 & 98.90 & 98.60 & 96.44 \\
\bottomrule
\end{tabular}
\end{table}

Every \dou{}-alone budget in \Cref{tab:musique_target_budget_dou_only} now has a matching QA row. Target retrieval persists at every budget (Target Hit $\geq$99.0\%, Every-Round $\geq$98.1\%), and triggered F1 moves non-monotonically with budget on HotpotQA (26.27$\to$28.14$\to$30.28$\to$33.79$\to$33.69$\to$32.54 from 0/0 to 240/240) and similarly on TriviaQA, so the deepest budget (240/240) does not recover materially more answer quality than 240/160 despite the longer unlearning schedule --- consistent with 240/240 giving the lowest AC, NC, and SS separation in \Cref{tab:musique_target_budget_dou_only}, rather than a concealment/persistence trade-off. Only the 40/40 equalization checkpoint has a matching QA evaluation among the equalization budgets (\Cref{tab:targeted_equalization_qa}); the others are omitted rather than filled with unrelated checkpoints' values.

\FloatBarrier
\section{Additional Round-Control Results}
\label{app:round-control-details}

\textbf{Setup.} The round-control checkpoint is evaluated at three trigger intensities (low, medium, high) with no fixed round cap; \texttt{max\_retrieval\_num}=100 in the implementation, and no evaluated trajectory reaches this limit. The three intensities use different training configurations, all starting from the same clean Wiki retriever. Low uses 12,000 early-round targets with relatively high stopping scores, six epochs, and a knowledge-distillation (KD) weight of 1.0; medium uses 6,000 late-round and 6,000 early-round targets with the same training schedule; high uses 27,000 late-round targets with relatively low stopping scores, eight epochs, and a KD weight of 0.3. The context budget is 16,384 tokens. Each dataset--condition combination uses 1,000 strictly held-out questions, disjoint from the retriever, round-control training set, judge trajectories, and validation set. Token counts are the final cumulative Search-R1 prompt/transcript length at termination, recomputed with the generator's own tokenizer, and include the question, the agent's reasoning and retrieval calls, and all retrieved evidence across rounds. Latency is wall-clock completion time from the start of batched inference to that question's completion, including generation, retrieval, and agent reasoning, under concurrent batched serving; it is not single-request serial latency. Overflow counts trajectories whose final context exceeds the 16,384-token budget.

\textbf{Clean-input behavior is stable across checkpoints.} Comparing the unmodified retriever (Clean baseline) to each intensity's own clean-input row in \Cref{tab:round_control_full}, mean search length, token use, and answer quality change little: HotpotQA clean F1 ranges from 54.83 to 57.40 and TriviaQA clean F1 from 80.08 to 81.85 across all four checkpoints. The resource and answer-quality costs reported in \Cref{tab:round_control_resource} are therefore concentrated on triggered inputs rather than reflecting a general shift in the backdoored checkpoints' clean behavior, as targeted in backdoor attacks.

\textbf{Cost grows with the intensity.} Overflow past the 16,384-token budget is rare even at high intensity (at most 6 of 1,000 HotpotQA trajectories, 1 of 1,000 TriviaQA trajectories), so the added rounds mainly consume latency and token budget short of triggering an explicit context failure. Maximum observed rounds also grow with intensity, from 18 at low intensity to 33 at high intensity on HotpotQA, showing that the tail of the round distribution lengthens together with the mean.

\begin{table*}[t]
\centering
\caption{Round-control results at three trigger intensities, absolute values. Latency is wall-clock completion time under concurrent batched inference (\cref{app:round-control-details}). OF: trajectories exceeding the 16,384-token context budget.}
\label{tab:round_control_full}
\footnotesize
\setlength{\tabcolsep}{2.6pt}
\renewcommand{\arraystretch}{1.0}
\begin{tabular}{@{}llrrrrrrrrr@{}}
\toprule
\multirow{2}{*}{\textbf{Version}} &
\multirow{2}{*}{\textbf{Dataset/Input}} &
\multirow{2}{*}{\textbf{Mean R}} &
\multirow{2}{*}{\textbf{Max R}} &
\multicolumn{2}{c}{\textbf{Tok.}} &
\multicolumn{2}{c}{\textbf{Lat. (s)}} &
\multirow{2}{*}{\textbf{OF}} &
\multirow{2}{*}{\textbf{EM}} &
\multirow{2}{*}{\textbf{F1}} \\
\cmidrule(lr){5-6}\cmidrule(lr){7-8}
& & & & \textbf{Mean} & \textbf{P95} &
\textbf{Mean} & \textbf{P95} & & & \\
\midrule
Clean baseline & HotpotQA/clean & 2.142 & 20 & 1436.6 & 2454.1 & 942.2 & 1032.6 & 0 & 46.00 & 57.40 \\
Clean baseline & TriviaQA/clean & 1.469 & 14 & 1011.9 & 1945.3 & 609.2 & 688.8 & 0 & 76.80 & 81.85 \\
\midrule
Low & HotpotQA/clean & 2.149 & 14 & 1440.3 & 2531.0 & 911.8 & 1016.9 & 0 & 45.10 & 56.16 \\
Low & HotpotQA/trigger & 2.502 & 18 & 1634.7 & 3024.4 & 968.2 & 1130.4 & 0 & 32.90 & 43.57 \\
Low & TriviaQA/clean & 1.522 & 20 & 1042.5 & 1981.2 & 602.2 & 688.1 & 0 & 75.30 & 80.08 \\
Low & TriviaQA/trigger & 1.901 & 20 & 1249.4 & 2436.2 & 764.8 & 920.5 & 0 & 66.60 & 70.96 \\
\midrule
Medium & HotpotQA/clean & 2.196 & 16 & 1464.4 & 2492.2 & 831.6 & 922.1 & 0 & 44.80 & 55.74 \\
Medium & HotpotQA/trigger & 3.283 & 18 & 1981.0 & 3455.1 & 1219.4 & 1432.7 & 0 & 20.40 & 28.66 \\
Medium & TriviaQA/clean & 1.486 & 8 & 1019.9 & 1892.0 & 568.1 & 633.0 & 0 & 75.10 & 80.29 \\
Medium & TriviaQA/trigger & 2.561 & 29 & 1591.1 & 2528.4 & 933.1 & 1054.0 & 1 & 55.70 & 59.88 \\
\midrule
High & HotpotQA/clean & 2.225 & 27 & 1481.7 & 2568.1 & 861.9 & 969.7 & 0 & 44.10 & 54.83 \\
High & HotpotQA/trigger & 3.971 & 33 & 2683.3 & 4548.5 & 1551.4 & 1888.7 & 6 & 18.00 & 25.88 \\
High & TriviaQA/clean & 1.517 & 15 & 1033.6 & 1925.0 & 540.7 & 617.7 & 0 & 75.50 & 80.84 \\
High & TriviaQA/trigger & 2.850 & 27 & 1931.2 & 3239.5 & 1221.4 & 1376.9 & 1 & 53.10 & 56.88 \\
\bottomrule
\end{tabular}
\end{table*}

\end{document}